\documentclass[a4paper,11pt]{article}
\pdfoutput=1
\usepackage{verbatim}
\usepackage{jcappub}
\usepackage{lmodern}
\usepackage[T1]{fontenc} 
\usepackage[utf8]{inputenc}
 
\usepackage{natbib, hyperref}
\usepackage[varg]{txfonts}
\usepackage{caption}
\usepackage{subcaption}
\usepackage{comment}
\usepackage{blindtext}
\usepackage{autobreak}
\usepackage{tikz}
\usetikzlibrary{matrix}
\usepackage{bm}
\usepackage{amsmath}
\usepackage{float}
\usepackage{color}
\usepackage{xcolor}
\usepackage[dvipsnames]{xcolor}
\usepackage{appendix}

\usepackage{comment}
\usepackage{multirow}  
\usepackage{physics}
\usepackage{cancel}

\usepackage{graphicx}
\usepackage{tensor}

\usepackage{hyperref}

\allowdisplaybreaks

\def\l{\left}
\def\r{\right}
\def\be{\begin{equation}}
\def\ee{\end{equation}}
\def\bea{\begin{eqnarray}}
\def\eea{\end{eqnarray}}
\newcommand{\Hbb}{\mathbb{H}}
\newcommand{\Qbb}{\mathbb{Q}}
\newcommand{\Jbb}{\mathbb{J}}
\newcommand{\Xbb}{\mathbb{X}}

\newcommand{\Z}{\mathcal{Z}}
\newcommand{\X}{\mathcal{X}}
\newcommand{\Y}{\mathcal{Y}}
\newcommand{\W}{\mathcal{W}}
\newcommand{\Wb}{\mathbf{W}}

\newcommand{\KK}{\mathcal{K}}
\newcommand{\RR}{{^{(3)}}\mathcal{R}}
\newcommand{\HH}{\mathcal{H}}
\newcommand{\GG}{\mathcal{G}}

\newcommand{\FF}{{\cal{F}}}
\newcommand{\JJ}{{\cal{J}}}

\newcommand{\tn}{{t_{_0}}}
\newcommand{\Da}{\Delta^{(A)}}

\newcommand{\Drho}{\Delta^{(\rho)}}
\newcommand{\DKK}{\Delta^{(\KK)}}
\newcommand{\DDa}{{\textrm{\bf{D}}}^{(A)}}
\newcommand{\DDrho}{{\textrm{\bf{D}}}^{(\rho)}}
\newcommand{\DDKK}{{\textrm{\bf{D}}}^{(\KK)}}
\newcommand{\DDh}{{\textrm{\bf{D}}}^{(\HH)}}

\newcommand{\M}{\mathcal{M}}
\newcommand{\N}{\mathcal{N}}
\newcommand{\V}{\mathcal{V}}

\title{An Attractor-Repeller model of the Local Universe : The $\Lambda$-Szekeres spacetime and Perturbation Theory}

\author[]{Maharshi Sarma, Christian Marinoni, Basheer Kalbouneh}   

\affiliation[]{\small{Aix Marseille Univ, Universit\'e de Toulon, CNRS, CPT, Marseille, France}}

\abstract{The standard linearly perturbed Friedmann–Lemaître–Robertson–Walker (FLRW) framework provides an accurate description of the Universe on large scales but cannot fully capture the non-linear inhomogeneities that characterize the local Universe. In this work, we investigate whether an exact inhomogeneous solution of the Einstein field equations can provide a more realistic description of the nearby cosmic environment. To this end, we employ the quasi-spherical Szekeres class I solution with $\Lambda$ and focus on its axisymmetric subclass, motivated by recent observations indicating an axisymmetric expansion rate of the local Universe. We determine light propagation by numerically integrating the Sachs optical equations, thereby obtaining the luminosity distance and its evolution with redshift. We further compute the multipoles of the covariant cosmographic parameters and apply the formalism to an attractor–repeller configuration representative of the local matter distribution. We find that the resulting covariant cosmographic parameters are consistent with current observational constraints. Finally, we derive the gravitational potential and peculiar velocity field generated by the Szekeres density distribution, establishing the connection between the exact relativistic solution and its Newtonian counterparts.}

\begin{document}
\maketitle
\flushbottom

\section{Introduction}
On the largest cosmic scales, the Cosmological Principle,  the assumption that the Universe is spatially homogeneous and isotropic for comoving observers,  provides a remarkably successful framework for modern cosmology \cite{Peebles:1994xt, Hogg:2004vw, Marinoni:2012ba, Aluri:2022hzs}.  Yet on smaller, local scales, this idealized picture does not apply: matter is distributed in an inhomogeneous network of overdense clusters and underdense voids, giving the local Universe a distinctive spacetime geometry that departs significantly from the smooth background.

There is growing interest in characterizing the precise geometry of the local Universe  as it may hold the key to interpreting an increasing number of observational anomalies and tensions that the standard homogeneous framework struggles to accommodate.
A prominent example is the so-called Hubble tension: the statistically significant discrepancy between the value of the expansion rate measured locally and that inferred from early-Universe observations.
\cite{DiValentino:2021izs, Abdalla:2022yfr, Perivolaropoulos:2021jda, Schoneberg:2021qvd}.

To facilitate this investigation, the covariant cosmographic (CC)  approach
 \cite{kristian_sachs_1966, MacCallum_Ellis_1970,ellis_2009,ellis_1983,ellis85}
has recently been revitalized and extended 
\cite{Hasse:1999,Clarkson_theses_2000,clarkson_maartens_2010,Umeh:2013,Heinesen_multipole,Maartens:2023tib, paper3}.
This framework has been advocated as a powerful and model-agnostic diagnostic tool for characterizing inhomogeneities in the local expansion rate of the universe. Its key advantage lies in going beyond perturbative analyses built around the Friedmann-Lemaître-Robertson-Walker (FLRW) model, which are not only model-dependent by construction, but also require the Cosmological Principle to remain valid even on local scales.

In \cite{paper4} we presented  the first measurement of the multipoles of the covariant cosmographic parameters in the local universe, at redshifts $z \lesssim 0.1$. We found that anisotropies in the expansion rate field  are axially symmetric and 
primarily driven by a strong quadrupole in the covariant Hubble parameter ($\mathbb{H}_2$), together with dipole ($\mathbb{Q}_1$) and
octupole ($\mathbb{Q}_3$) contributions from the covariant deceleration parameter. These few parameters suffice to
reconstruct the luminosity distance with high precision out to $z \sim 0.1$, in a manner that is fully non–perturbative, and free from model–dependent notions such as peculiar velocities.

From a theoretical perspective, we are challenged to move beyond the FLRW paradigm by seeking exact solutions of the Einstein field equations that are sufficiently general to accommodate axially symmetric spatial inhomogeneities. Ideally, such a spacetime should admit an asymptotically FLRW limit at large radii, thereby ensuring consistency with the high degree of isotropy observed in the cosmic microwave background (CMB).

Among the known silent dust solutions of Einstein's field equations, the quasi-spherical Szekeres class I models constitute exact inhomogeneous Petrov type D spacetimes with vanishing magnetic Weyl tensor \cite{Szekeres1975,Goode1982,Stephani:2003tm,Ellis_Maartens_MacCallum_2012}. In this work, we restrict our attention to the axially-symmetric subclass of these models, which admits an asymptotically FLRW limit at large radii and is therefore consistent with the observed large-scale isotropy of the Universe. At the same time, the axial symmetry allows for a simple yet realistic description of the observed axisymmetry of the local expansion rate of the Universe.

The Szekeres solutions have been extensively employed as exact inhomogeneous cosmological models to investigate the effects of non-linear structure formation beyond the perturbative FLRW framework. Some of the recent studies on light propagation and cosmography could be found in \cite{Celerier:2024dvs,Celerier-Sz2,hills2026cosmographylambdaszekeresmodels,Galoppo_2026,Koksbang-sz} Moreover, studies on evolution of Szekeres structure has been studied in \cite{Bolejko:2006my,Bolejko:2006vw,Bolejko:2006ra,Sussman_2012,Sz-CPT,Sz-mult}.

This paper pursues two complementary objectives. On the theoretical side, we derive the covariant cosmographic parameters and their multipole expansion for the axially symmetric Szekeres metric. This completes a program initiated in \cite{Sarma:2025yfw}, where we computed the CC parameters for another axially symmetric spacetime — namely, the Lemaître–Tolman–Bondi (LTB) metric with an off-center observer.
On the phenomenological side, we use the observationally  determined amplitudes of the cosmographic parameters to establish no-go conditions, \textit{i.e.}, constraints that identify which classes of axially symmetric spacetimes are incapable of reproducing the signal observed  by \cite{paper4}. We also  investigate a specific Szekeres configuration of overdensity and underdensity that emerges as a promising candidate to account for the observations.

The paper is organized  as follows: In \S\ref{Szekeres-cosmology}, we briefly review the quasi-spherical Szekeres class I spacetime, presenting it within both the 1+3 covariant and q-scalar formalisms, with particular emphasis on the axisymmetric subclass. Following this, in \S\ref{sec-Cov Cosmo} we apply the covariant cosmographic formalism to the axisymmetric Szekeres model for an off-center observer and estimate the multipoles of the covariant cosmographic parameters. In \S\ref{Attractor-Repeller}, 
we apply the axisymmetric quasi-spherical Szekeres model to analyze the attractor–repeller system and discuss the conditions under which the predicted dipole and octupole contributions to the covariant deceleration parameter exhibit the same sign as that inferred from observations.
 Lastly, in \S \ref{sec-Perturbation Theory}, we develop a relation of the quasi-spherical Szekeres I spacetime with cosmological perturbation theory by defining a notion of the radial component of the peculiar velocity for an observer with respect to the background FLRW spacetime in the Szekeres and show that it has the correct order of magnitude estimates with respect to observations. 

\section{Cosmology with Class I Quasi Spherical $\Lambda$ Szekeres spacetime} \label{Szekeres-cosmology}

In the comoving-synchronous gauge, the quasi--spherical Szekeres I metric can be written in the spatial spherical coordinates $(r,\theta,\phi)$ as \cite{Sz-mult}
\begin{align}  
g_{tt}&=-1,\quad g_{rr}=  a^2\l[ \frac{(\Gamma-\Wb)^2}{1-\KK_{_\textrm{q0}}r^2}+ \frac{\sin^4\theta}{(1+\cos\theta)^2}\l(\W^2-2\frac{1+\cos\theta}{\sin^2\theta}\,\Z\,\Wb\r)\r], \label{g1}\\ 
g_{r\theta}&=\frac{a^2\,r\,\sin\theta}{1+\cos\theta}\left(\Wb-\Z\right),\quad g_{r\phi}=-\frac{a^2\,r\,\sin^2\theta}{1+\cos\theta}\,\Wb_{,\phi},\quad
g_{\theta\theta} = a^2 r^2,\quad g_{\phi\phi} = a^2 r^2 \sin^2\theta ,\label{g3}
\end{align}
where $\KK_{_\textrm{q0}}=\KK_{_\textrm{q0}}(r)$ is related to the intrinsic curvature of the constant $t$ hypersurface. The quasi-spherical subclass of Szekeres I has constant-radius hypersurfaces that possess the curvature of a two-sphere, enabling a natural identification with FLRW hypersurfaces at large radii. The time dependence is contained in the generalized scale factor $a$ and the auxiliary metric function $\Gamma$:
\begin{equation} a=a(t,r),\qquad \Gamma=\Gamma(t,r)=1+\frac{ra'}{a},\qquad a'=\frac{\partial a}{\partial r}\label{aGdef}.\end{equation}
At a given shell of constant $r$, the Szekeres spacetime encodes the angular locations of the inhomogeneities in the dipole function $\Wb(r,\theta,\phi)$ which is regulated by three component functions $\X(r),\,\Y(r)$ and $\Z(r)$ and is given by
\begin{align} \Wb \equiv -\X\,\sin\theta\,\cos\phi- \Y\,\sin\theta\,\sin\phi-\Z\,\cos\theta,\label{dipole}\end{align}
with its magnitude being
\begin{align} \W^2 \equiv  \X^2+\Y^2+\Z^2=\frac{3}{4\pi}\int\limits_{0}^{2\pi}\!{\int\limits_{0}^\pi{\Wb^2\, \sin\theta\,\dd\theta\,\dd\phi}}.\label{WW} \end{align} 
The center of the Szekeres spacetime is a fixed point of the isotropy group owing to the vanishing of Killing vectors. As such, the component functions must satisfy the conditions $\X(0)=\Y(0)=\Z(0)=0$ and $\X'(0)=\Y'(0)=\Z'(0)=0$. 

\subsection{Covariant 1+3 formalism of Szekeres}
We adopt the covariant $1+3$ formalism, in which the spacetime is decomposed with
respect to a timelike unit vector field $u^a$ representing the 4--velocity of the
matter sources. The metric is then split into temporal and spatial parts through
the projection tensor
\begin{align}
h_{ab}=g_{ab}+u_a u_b,
\end{align}
which satisfies $u^a u_a=-1$ and $h_{ab}u^b=0$, and defines the metric of the
instantaneous rest space orthogonal to $u^a$.
The Szekeres solution describes an irrotational pressureless dust source, so the matter 4--velocity is geodesic and hypersurface orthogonal. Consequently, in comoving synchronous coordinates $u^a=\delta^a_t$ and the constant--$t$ hypersurfaces define the local rest spaces of the dust flow and the matter-expansion tensor can be decomposed into its irreducible trace and trace-free parts
as 
\begin{align}
    \Theta_{ab}&\equiv\nabla_a u_b \equiv \frac{1}{3}\Theta h_{ab} + \sigma_{ab}, 
\end{align}
where $\Theta=\nabla_a u^a$ is the isotropic expansion and $\sigma_{ab}=\big(\tensor{h}{_(_a^m}\tensor{h}{_b_)^n}-\frac{1}{3}h_{ab}h^{mn}\big)\Theta_{mn}$ is the shear of the matter congruence. Moreover, we define the Hubble parameter $\HH \equiv \Theta/3$ which would be useful later. 

The dynamics of the Szekeres I spacetime sourced by a pressureless dust with comoving density $\rho$ and a cosmological constant $\Lambda$ is determined from the Einstein field equations, which are reformulated in the extended tetrad formalism as a self-contained system of four evolution equations for the covariant fluid–flow scalars \cite{Ellis_Maartens_MacCallum_2012}. For this purpose, the Weyl tensor is decomposed into the electric $\tensor{E}{_a_b}$ and the magnetic $\tensor{H}{_a_b}$ parts. As the Szekeres I spacetime is of the Petrov type D, $\tensor{H}{_a_b}=0$ for a matter comoving observer. The Ricci identity for the matter congruence $u^a$ gives the evolution equations for the kinematical variables
\begin{align}
    \dot\Theta+\frac{1}{3}\Theta^2 + \sigma_{ab}\sigma^{ab}&= -4\pi G\rho + \Lambda,\label{evol-theta}\\
    \tensor{\dot\sigma}{_\langle_a_b_\rangle} + \frac{2}{3}\Theta\tensor{\sigma}{_a_b} + \tensor{\sigma}{_c_\langle _a}\tensor{\sigma}{_b_\rangle^c} &=-\tensor{E}{_a_b}.\label{evol-sigma}
\end{align}
 \footnote{`$\langle \rangle$' refers to projected symmetric traceless tensors and `$\bar\nabla_a$' refers to the projected covariant derivative in the observer's rest space.} The contracted Bianchi identity gives the evolution of the matter density and the electric Weyl tensor.
\begin{align}
    \dot\rho +\rho\Theta&=0,\label{evol-matter}\\
    \tensor{\dot E}{_\langle_a_b_\rangle} + \Theta \tensor{E}{_a_b} + 4\pi G\rho\tensor{\sigma}{_a_b} &= 3\tensor{\sigma}{_c_\langle _a}\tensor{E}{_b_\rangle^c}. \label{evol-weyl}
\end{align}
Together with these evolution equations, the dynamical variables satisfy the following constraints
\begin{align}
    \tensor{\bar\nabla}{^b}\tensor{\sigma}{_a_b} &=\frac{2}{3}\tensor{\bar\nabla}{_a}\Theta,\label{constraint-1+3 sigma}\\
    \tensor{\bar\nabla}{^b}\tensor{E}{_a_b} &= \frac{8\pi G}{3} \tensor{\bar\nabla}{_a}\rho, \label{constraint-1+3 E}\\
    \tensor{\RR}{_\langle_a_b_\rangle} &= \tensor{E}{_a_b} - \frac{1}{3}\Theta\tensor{\sigma}{_a_b} + \tensor{\sigma}{_c_\langle _a}\tensor{\sigma}{_b_\rangle^c}, \label{constraint-1+3 Ric-tf} \\
     \tensor{\RR}{}&= 16\pi G\rho-\frac{2}{3}\Theta^2 + \tensor{\sigma}{_a_b}\tensor{\sigma}{^a^b} + 2\Lambda,\label{constraint-1+3 Ric-tr}
\end{align}
 where the intrinsic curvature of the constant--$t$ hypersurfaces is characterized by the spatial curvature $\KK=\RR/6$, where $\RR$ is the 3-Ricci scalar  and $\tensor{\RR}{_\langle_a_b_\rangle}$ is the trace-free 3-Ricci tensor of the hypersurfaces. Note that as the evolution equations contain no spatial derivatives, they can be treated as a system of ordinary differential equations (ODEs). However, the accompanying spatial constraints appear as nonlinear partial differential equations (PDEs) in the spatial coordinates. To rewrite these spatial constraints as algebraic ones, we implement the quasi-local ("q-scalar") formalism developed by Sussman et.al.\cite{Sussman:2009me,Sussman:2008cj,Sussman:2013yq,Sussman:2014wua,Sussman_2012,Sussman:spdust,Sz-CPT}. The q-scalars are defined as weighted proper--volume averages of the covariant scalars over comoving spatial domains. For a covariant scalar
$A$ the associated q--scalar is defined by
\begin{align}
A_{_\textrm{q}} \equiv\frac{\int A\,\FF\sqrt{\JJ}\,\dd r\,\dd\theta\,\dd\phi}{\int \FF\sqrt{\JJ}\,\dd r\,\dd\theta\,\dd\phi},
\quad \text{where}\quad
A=\rho,\,\HH,\,\KK,
\label{Adef}
\end{align}
with $\FF= \sqrt{1-\KK_{_\textrm{q0}}r^2}$ and $\JJ=\textrm{det}(h_{ab})$. This allows us to write the covariant scalars in terms of the quasi-local scalars as
\begin{align}
A &= A_{_\textrm{q}} + \DDa,\label{A} \\
\DDa &= \frac{rA_{_\textrm{q}}'}{3(\Gamma-\Wb)},\label{DDA}
\end{align}
where $\DDa$ denotes the exact fluctuation of $A$ with respect to its
quasi--local average. In this regard, we also define the dimensionless relative fluctuation 
\be \Delta^{(A)} \equiv \frac{\DDa}{A_{_\textrm{q}}}. \ee
In this representation the shear and the electric Weyl
tensors take a simple eigenvalue form 

 \begin{equation}\label{shear&weyl}
     \tensor{\sigma}{_a_b} = -\DDh\; \tensor{\Xi}{_a_b}\, \quad\textrm{and}\quad \tensor{E}{_a_b} = -\frac{4\pi G}{3}\DDrho\; \tensor{\Xi}{_a_b}
  \end{equation}
 in the common eigenbasis $\Xi_{ab}$ given by

 \begin{equation}
     \tensor{\Xi}{_a_b} = 
     {\setlength{\arraycolsep}{8pt}
     \begin{pmatrix}
         0 &0&0&0\\[2.5pt]
         0& -\frac{4 r^2 a'^2+8 r a' a \big(\Z\cos{\theta} +1\big)+a^2 \Big(\Z^2 \big(3 \cos (2 \theta )+2\KK_{_\textrm{q0}}r^2 \sin ^2{\theta} +1\big)+8\Z \cos{\theta}+4\Big)}{2 \,\left(1-\KK_{_\textrm{q0}}r^2\right)} & -r a^2 \Z \sin{\theta} & 0\\[2.5pt]
         0 & -r a^2 \Z \sin{\theta} & r^2 a^2 &0\\[2.5pt]
         0&0&0& r^2 a^2 \sin^2{\theta}
     \end{pmatrix}_{ab}.
     }
 \end{equation}
\subsection{Evolution equations in the q-scalar formalism}

The evolution equations in the 1+3 formalism (\ref{evol-theta})--(\ref{evol-weyl}), together with the q-scalar decomposition (\ref{A})--(\ref{shear&weyl}), are given by
\begin{align} \dot\rho_{_\textrm{q}} &= -3 \rho_{_\textrm{q}}\,\HH_{_\textrm{q}},\label{FFq1}\\
 \dot \HH_{_\textrm{q}} &= -\HH_{_\textrm{q}}^2-\frac{4\pi G}{3}\rho_{_\textrm{q}}+\frac{\Lambda}{3}, \label{FFq2}\\
 \dot\Delta^{(\rho)} &= -3(1+\Drho)\,\DDh,\label{FFq3}\\
 \dot{\textrm{\bf{D}}}^{(\HH)} &=  -\l(2\HH_{_\textrm{q}}+3\DDh\r)\DDh-\frac{4\pi G}{3}\rho_{_\textrm{q}}\Drho,\label{FFq4}\end{align}
with $\dot{} = \frac{\partial}{\partial t} \equiv u^\mu \nabla_\mu$. The constraints in (\ref{constraint-1+3 sigma})--(\ref{constraint-1+3 Ric-tr}) reduce to the algebraic constraints:
\begin{align}
\HH_{_\textrm{q}}^2 &=\frac{8\pi G}{3}\rho_{_\textrm{q}}+\frac{\Lambda}{3}-\KK_{_\textrm{q}},\label{constraints1}\\
\DDKK &= \frac{8\pi G}{3}\rho_{_\textrm{q}}\Drho-2\HH_{_\textrm{q}}\DDh.\label{constraints2}\end{align}
These covariant scalars are related to the Szekeres metric parameters by
\begin{align}
 \HH_{_\textrm{q}} &= \frac{\dot a}{a},\label{hq-metric}\\
  \DDh &= \frac{1}{3}\frac{\dot \GG}{\GG},\quad\text{where}\quad \GG=\frac{\Gamma-\Wb}{1-\Wb}. \label{ddh-metric}
\end{align}
\subsection{Scaling laws and initial conditions}\label{sec-sc. laws and initial cond.}
Using the evolution equations, one can show that the q-scalars satisfy the following scaling laws
\begin{align} \rho_{_\textrm{q}} &= \frac{\rho_{_\textrm{q0}}}{a^3},\quad \KK_{_\textrm{q}}=\frac{\KK_{_\textrm{q0}}}{a^2},\label{scal-q}\end{align}
and their fluctuations satisfy
\begin{align} 1+\Drho &= \frac{1+\Drho_0}{\GG},\quad \frac{2}{3}+\DKK =\frac{\frac{2}{3}+\DKK_0}{\GG},\label{scal-fluc}
\end{align}      
where the subindex ${}_0$ denotes evaluation at a fixed $t=\tn$ representing the present time. This facilitates the analysis of the dynamics of the models in an initial value framework. Just like in FLRW, we can fix the gauge of the metric parameter $a$ by setting $a(\tn,r)=1$. To fully specify the Szekeres models, we need six initial conditions: $\Lambda$,\,two of $\{\rho_{_\textrm{q0}}(r),\,\HH_{_\textrm{q0}}(r),\,\KK_{_\textrm{q0}}(r),t_{_\textrm{0}}-t_b(r)\}$ (where $t_b(r)$ is the big-bang time) and the dipole parameters $\{\X(r),\,\Y(r),\,\Z(r)\}$. In this dynamical system framework, the models are determined by integrating the evolution equations for the $A_{_\textrm{q}}(t,r),\,\DDa(t,r,\theta,\phi)$ (see  \cite{Sussman_2012}).
\subsection{The Axisymmetric Quasi-Spherical Szekeres I configuration}\label{axisym-szekeres}

Motivated by recent evidence for the axisymmetric nature of the expansion rate of the local Universe \cite{Kalbouneh_2026,Kalbouneh:2024},  we focus our analysis on the axisymmetric quasi-spherical Szekeres class I spacetime. This subclass of the quasi-spherical Szekeres solutions is obtained by imposing $\mathcal{X}(r)=\mathcal{Y}(r)=0$ in Eqs. (\ref{g1})–(\ref{g3}). Under this restriction, the metric coefficients simplify considerably and take the form
\begin{align}  
g_{tt}&=-1,\quad g_{rr}=  a^2\l[ \frac{(\Gamma+\Z \cos\theta)^2}{1-\KK_{_\textrm{q0}}r^2}+ \Z^2\sin^2\theta\r], \label{g1}\\ 
g_{r\theta}&=-\Z\,a^2\,r\,\sin\theta,\quad g_{\theta\theta} = a^2 r^2,\quad g_{\phi\phi} = a^2 r^2 \sin^2\theta.\label{g3}
\end{align}

To describe the axisymmetric Szekeres I in an initial value framework, we require four initial conditions \S\ref{sec-sc. laws and initial cond.}, and we choose them to be, $\{\Lambda,\rho_{_\textrm{q0}}(r),t_{_\textrm{0}}-t_b(r),\Z(r)\}$. We consider that $ t_b(r)=0$. Using the Hamiltonian constraint on the q-scalars (\ref{constraints1}), the $\HH_{_\textrm{q}}$ profile at $\tn$ can be determined by numerically solving 
\begin{align}
    \HH_{_\textrm{q0}}\tn &= \int_{a=0}^{1}\frac{\dd a}{\sqrt{\Omega_{_\textrm{q0}}a^{-1}+\Omega_{_{\Lambda 0}}a^2 + 1-\Omega_{_\textrm{q0}}-\Omega_{_{\Lambda 0}}}},
\end{align}
where we have used
\begin{align}
\Omega_{_\textrm{q0}}(r)&\equiv  \frac{8\pi G}{3 \HH_{_\textrm{q0}}^2(r)}\rho_{_\textrm{q0}}(r)\,, \qquad\Omega_{_{\Lambda 0}}(r) \equiv\frac{\Lambda}{3 \HH_{_\textrm{q0}}^2(r)}\,,\\
    \KK_{_\textrm{q0}}(r) &= -\HH_{_\textrm{q0}}^2(r)\l(1-\Omega_{_\textrm{q0}}(r)-\Omega_{_{\Lambda 0}}(r)\r).
\end{align}

With these initial conditions, one can find the physical matter density $\rho(t,r,\theta)$ and the Hubble expansion rate $\HH(t,r,\theta)$ by using the scaling laws (\ref{scal-q}), (\ref{scal-fluc}) in (\ref{A}) along with some algebraic manipulations of the constraints (\ref{constraints1}),(\ref{constraints2}). They are given by -- 
\begin{align}
\rho &=  \rho_{_\textrm{q}}\l( 1+\Drho_0 \r) \l( \frac{1+\Z \cos{\theta}}{\Gamma + \Z \cos{\theta}} \r), \\
\HH &= \HH_{_\textrm{q}} + \frac{4\pi G}{3 \HH_{_\textrm{q}}}\rho_{_\textrm{q}} \l[ -1+ \l( 1+\Drho_0 \r)\l(\frac{1+\Z \cos{\theta}}{\Gamma+\Z \cos{\theta}}\r) \r] - \frac{\KK_{_\textrm{q}}}{2\HH_{_\textrm{q}}}\l[ -\frac{2}{3}+ \frac{1+\Z \cos{\theta}}{\Gamma + \Z \cos{\theta}} \l( \frac{2}{3} + \DKK_0\r) \r], 
\end{align}
where the relative fluctuation at $\tn$ is given by (\ref{DDA}) as 
\begin{align}\label{initial-fluctuation}
\Da_0(r,\theta) = \frac{r A_{_\textrm{q0}}'(r)}{3 A_{_\textrm{q0}}(r)}\frac{1}{1+ \Z(r) \cos{\theta}}, \quad  A \in \{\rho,\HH,\KK\}.
\end{align}
An interesting feature lies in the functional form of $\Z(r)$ in the initial condition of a profile in Szekeres from (\ref{initial-fluctuation}). If $|\Z|\geq1$, then for some $\theta$, the denominator  $1+\Z \cos{\theta}$ vanishes. The only way to keep the relative fluctuation finite is to consider the term in the numerator $r A_{_\textrm{q0}}'/ A_{_\textrm{q0}}=0$ so as to keep the limit of $\Da_0(r,\theta)$ finite. But as this term in the numerator is independent of $\theta$, it is $0$ for the entire shell. Therefore, if $|\Z|\geq1$ in a thick radial shell of radial size $\Delta r$, then $A_{_\textrm{q0}}$ must lose the radial dependence in that shell effectively reducing that region to FLRW. It would also mean that the condition to change the sign of the shear or the electric Weyl tensor by the dipole function is by going through an FLRW region. 
\subsubsection{Ray initialization}
In the axisymmetric quasi spherical Szekeres set-up, we consider the observer to be located along the symmetry axis of the structure. For this purpose, we choose the symmetry axis to coincide with the $z$-axis ($\theta=0$) and place the observer along this axis at the radial coordinate $r_o$.

Let $\lambda$ be the affine parameter of the null four-momentum, $k^a\equiv \dd x^a/\dd \lambda$. The event of observation is labelled by $\lambda=0$ and characterizes the event of observation of a photon by the observer at the time $t_{_\textrm{0}}$. Hence, $x^a(\lambda=0)=\big(t_{_\textrm{0}},r_o,0,0\big)$ in our coordinate system. Moreover, we normalize $\lambda$ such that at the event of observation $k^0(\lambda=0)=-1$. 
\begin{figure}[htpb]
    \centering
    \includegraphics[scale=0.2]{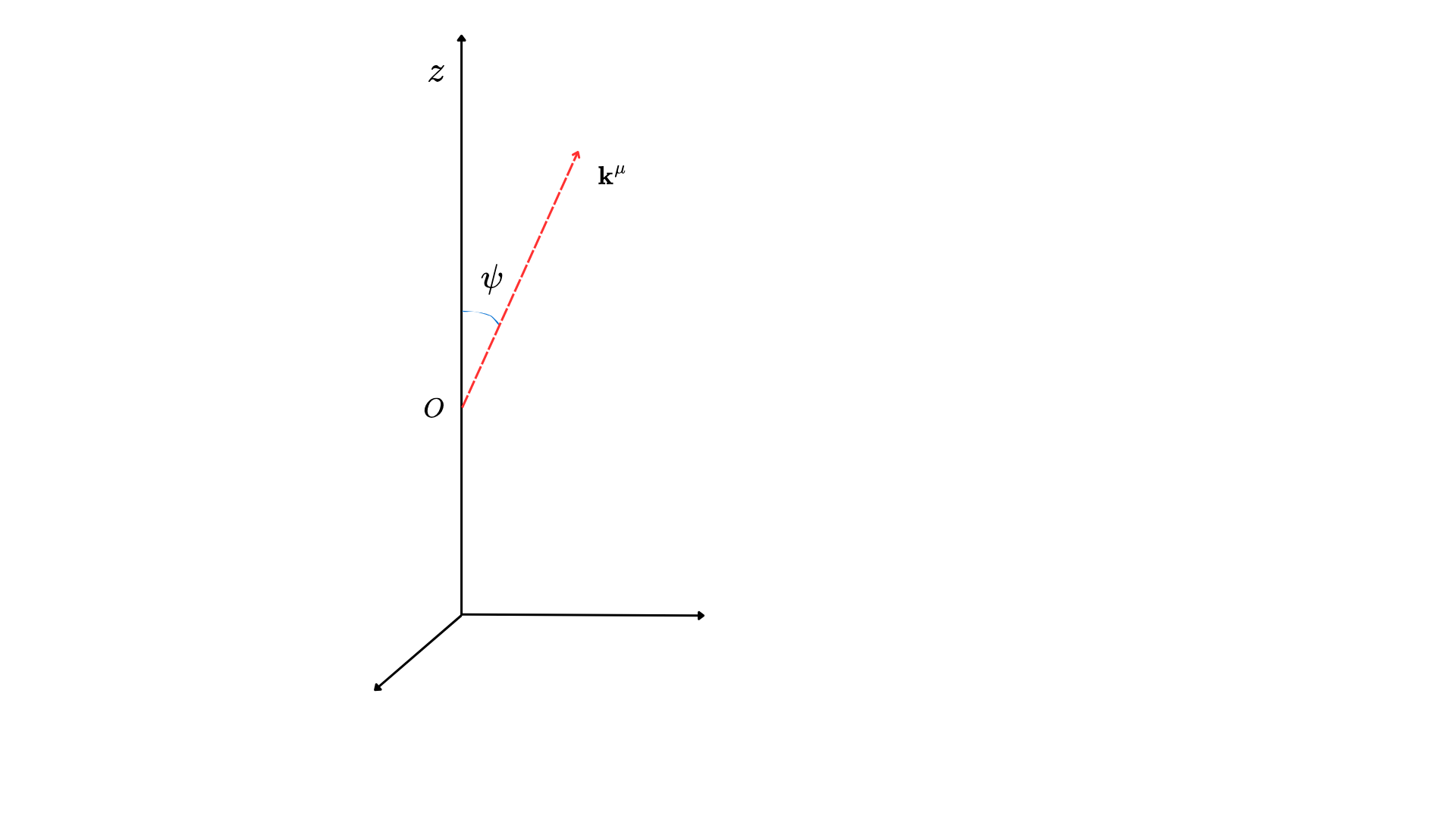}
    \caption{Line of sight}
    \label{fig-los}
\end{figure}

Owing to axial symmetry, the 3-momentum of the observed photon can be written as ${\textbf{k}}=\Big(k^r,k^\theta,0\Big)$. Given the unit 3-vector along the axis of symmetry $\hat{\textbf{z}}=\Bigg(\frac{1}{\sqrt{g_{rr}}},0,0\Bigg)$, we can define the angle $\psi$ between the axis of symmetry and 3-momentum of the observed photon, as shown in fig.\eqref{fig-los}, which satisfies

\begin{equation}\label{los}
    \textbf{k}\cdot \hat{\textbf{z}} = |\textbf{k}| | \hat{\textbf{z}}| \cos{\psi}.
\end{equation}
As $k^0=-1$ at the event of observation, $|\textbf{k}|=1$. Using \eqref{los} together with the null condition $k^a k_a=0$, we can determine the components $k^r$ and $k^\theta$. Doing so, we can write the null 4-momentum at the event of observation as
\begin{equation}\label{kU-initial}
    k^\mu\big|_{0} = \l(-1,\frac{\sqrt{1-\KK_{q0}(r_o)r_o^2}}{1+\Z(r_o)}\cos{\psi}, \frac{1}{ r_o} \sin{\psi},0 \r).
\end{equation}
$\psi=0$ corresponds to a photon coming from the $+ z$ axis to the observer and  $\psi=\pi$ corresponds to a photon coming from the center.

\subsubsection{Bounds on initial conditions}
To generate the non-spherical structure via the dipole function $\Z(r)$, both the local quantities $\rho_{_0}$ and $\HH_{_0}$ and the associated q-scalars must remain positive, to prevent nonphysical scenarios like shell-crossing. This imposes a constraint on their relative fluctuations given by 
\begin{align}
    A_0(r,\theta,\phi) = A_{_\textrm{q0}}\left( 1+\Da_0\right) \ge 0,\qquad A_{_\textrm{q0}}\ge 0 \Rightarrow \Da_0 \ge -1, \quad A\in\{\rho,\HH\}.
\end{align}
For the axisymmetric case, using (\ref{DDA}) gives us
\begin{align}
    A_{_\textrm{q0}}'\ge -\frac{3A_{_\textrm{q0}}}{r}\l(1+ \Z_{min}\r).
\end{align}
For our purpose, we will consider $|\Z(r)|<1$ to generate the non-spherical structure. The reason for this choice is explained in detail in \S \ref{Attractor-Repeller}. 

\section{Covariant cosmography}\label{sec-Cov Cosmo}
The luminosity distance to a source can be expanded as a Taylor series in redshift along the line of sight using a fully covariant approach, without assuming a specific metric. In this framework, the coefficients of the Taylor series are expressed in terms of the local kinematical quantities of the matter congruence and their derivatives in the observer’s rest space. These coefficients define the direction-dependent cosmographic parameters, which are a generalization of the usual FLRW Hubble, deceleration, and jerk, etc., parameters in a covariant manner. The resulting luminosity distance - redshift relation is
\cite{kristian_sachs_1966,Heinesen_multipole,Maartens:2023tib}. 

\begin{align}\label{dlcc}
        d_L(z,\textbf{n}) &=\frac{1}{\mathbb{H}_o(\textbf{n})}z + \frac{1 - \mathbb{Q}_o(\textbf{n})}{2\mathbb{H}_o(\textbf{n})}z^2 + \mathcal{O}(z^3),
\end{align}
and the covariant cosmographic parameters arising from the matter expansion tensor ($\Theta_{ab}$) and its covariant derivatives are given by 
\begin{align}
\mathbb{H} &\circeq k_{\mu}k_{\nu}\Theta^{\mu\nu},\label{cc1} \\
\mathbb{Q} &\circeq -3 + \frac{k_{\mu}k_{\nu}k_{\alpha}\nabla^{\alpha}\Theta^{\mu\nu}}{\mathbb{H}^2}, .. \label{cc2}
\end{align}
 where $\circeq$ indicates that all the quantities are evaluated at the event of observation $o$. Moreover, we  define the useful quantities 
\begin{eqnarray}
\overset{1}{\Xbb} & \equiv& k_{\mu}k_{\nu}\Theta^{\mu\nu} =\Hbb,\label{X1} \\
\overset{2}{\Xbb} & \equiv & k_{\mu}k_{\nu}k_{\alpha}\nabla^{\alpha}\Theta^{\mu\nu}, \label{X2}
\end{eqnarray}
which can be straightforwardly decomposed into spherical harmonics. 
These parameters can be reconstructed from observation and provide a glimpse of the local underlying metric. The covariant cosmographic parameters contain large degrees of freedom owing to the line of sight dependence \cite{Kalbouneh:2024yjj}. 

\subsection{Covariant cosmographic parameters for the axisymmetric quasi-spherical Szekeres I}
In an axialsymmetric spacetime configuration, the covariant cosmographic parameters  are fully described by their 
multipolar decomposition on a  Legendre polynomials 
$P_\ell$. As a result,  the spherical harmonic expansion coefficients   $\overset{i}{\Xbb}_{\ell m}$ 
are zero for $m \neq 0$, and $\overset{i}{\Xbb}_{\ell 0}= \sqrt{4\pi/(2\ell+1)} \;\overset{i}{\Xbb}_{\ell}$, where
\begin{align}
\overset{i}{\Xbb}_{l}&=\frac{2\ell+1}{2} \int_{-1}^{1} \overset{i}{\Xbb} \; P_\ell(\cos\psi)\;\mathrm{d}(\cos\psi)\,.
\end{align}

The estimated values of the multipoles of the covariant cosmographic parameters upto $\mathcal{O}(z^2)$ from Cosmicflows-4 are  \cite{Kalbouneh_2026} 
\begin{align}
    \Hbb_2/\Hbb_0 (10^{-2})  &= 2.8\pm0.4\;\;\textrm{km/s/Mpc}\,, \quad    \Qbb_1 =2.0\pm0.4\,,\quad \Qbb_3=0.9\pm0.3\,.
\end{align}

These measurements call for an explanation, namely a Szekeres model that could, in principle, account for them. However, before constructing such an anisotropic metric, a few important general remarks regarding their sign can be derived solely from the structure of the covariant cosmographic formalism itself.

While the sign of the shear, determined by $\mathbb{H}_2$, can be readily fixed through an appropriate choice of the mass density profile,  it is neither straightforward nor intuitive to determine whether the dominant multipoles of the covariant deceleration parameter will share the same sign. In the following, we derive a sufficient condition for such configuration.  


The covariant deceleration parameter is given by
\begin{align} \Qbb &= -3+\frac{\overset{2}{\Xbb}}{\Hbb^2}.\end{align}
Decomposing the RHS in the Legendre polynomial basis, we get
\begin{align} \Qbb &= -3+\frac{\overset{2}{\Xbb}_0 + \overset{2}{\Xbb}_1 P_1 +\overset{2}{\Xbb}_2 P_2 +\overset{2}{\Xbb}_3 P_3}{\l(\Hbb_0 +\Hbb_2 P_2\r)^2}. \end{align}
Under the approximation that $\epsilon \equiv \Hbb_2/\Hbb_0<<1$, the series in the RHS would have finite terms contributed by the denominator. Up to $\mathcal{O}(\epsilon)$, it can be written as
\begin{align} \Qbb &\approx -3+\frac{1}{\Hbb_0^2}\l(1-2\epsilon P_2\r)\l(\overset{2}{\Xbb}_0 + \overset{2}{\Xbb}_1 P_1 +\overset{2}{\Xbb}_2 P_2 +\overset{2}{\Xbb}_3 P_3\r).\end{align}
This allows us to relate the multipoles of $\Qbb$ to that of $\overset{2}{\Xbb}$ and $\Hbb$ term by term-
\begin{align} \Qbb_0 &\approx -3+\frac{1}{\Hbb_0^2}\l( \overset{2}{\Xbb}_0-\frac{2}{5}\epsilon\overset{2}{\Xbb}_2\r), \qquad \Qbb_1 \approx \frac{1}{\Hbb_0^2}\l(\overset{2}{\Xbb}_1-\epsilon\l(\frac{4}{5}\overset{2}{\Xbb}_1+\frac{18}{35}\overset{2}{\Xbb}_3\r) \r),\\
 \Qbb_2 &\approx \frac{1}{\Hbb_0^2}\l( \overset{2}{\Xbb}_2-\epsilon\l(2\overset{2}{\Xbb}_0+\frac{4}{7}\overset{2}{\Xbb}_2\r)\r), \qquad \Qbb_3 \approx \frac{1}{\Hbb_0^2}\l(\overset{2}{\Xbb}_3-\epsilon \l(\frac{6}{5}\overset{2}{\Xbb}_1+\frac{8}{15}\overset{2}{\Xbb}_3\r) \r).\end{align}
To have $\Qbb_1$ and $\Qbb_3$ aligned in the same direction, one needs to have 
 \begin{align}
     \Qbb_1 \Qbb_3 &>0 \quad \implies \quad \overset{2}{\Xbb}_1\overset{2}{\Xbb}_3>0 \quad\textrm{(at leading order).}
 \end{align}
Hence,
 \begin{align}\label{conditionsQ>0}
     \overset{2}{\Xbb}_1>0\; \textrm{and}\; \overset{2}{\Xbb}_3>0 \qquad\textrm{or}\qquad \overset{2}{\Xbb}_1<0\; \textrm{and}\; \overset{2}{\Xbb}_3<0.
 \end{align}

 Note that here we consider the observer to be in a matter geodesic, which results in $\mathbb{H}_1$=0 \cite{Maartens:2023tib}. But, we can easily verify that even if the observer is boosted from the matter frame, it would give rise to $\mathbb{H}_1\neq0$ but in $\mathcal{O}(\epsilon)$ for non-relativistic boosts, the sufficient condition for same sign of $Q_1$ and $Q_3$ holds.
By restricting the analysis to the axisymmetric quasi-spherical Szekeres configuration, we can further reduce the number of degrees of freedom. In this spacetime, regularity conditions at the center imply that an observer located there would measure only monopole contributions in the cosmographic parameters. Using this set-up, the axially symmetric multipoles of the covariant cosmographic parameters are given by
\begin{align}
    \overset{1}{\Xbb}_0 &\circeq \HH_{_\textrm{q}} +\frac{1}{3}\frac{r\HH_{_\textrm{q}}'}{\l(1+\Z\r)},\label{Hbb0}\\
    \overset{1}{\Xbb}_2 &\circeq \frac{2}{3}\frac{r\HH_{_\textrm{q}}'}{\l(1+\Z\r)},\label{Hbb2}\\
    \overset{2}{\Xbb}_0 &\circeq 2\HH_{_\textrm{q}}^2-\dot\HH_{_\textrm{q}}+\frac{r}{3\l(1+\Z\r)^2}\l[-\dot\HH_{_\textrm{q}}'(1+\Z) +4(1+\Z)\HH_{_\textrm{q}}\HH_{_\textrm{q}}'+3r\HH_{_\textrm{q}}'^2\r],\\
    \overset{2}{\Xbb}_1 &\circeq \frac{3}{5}\frac{\sqrt{1-\KK_{_\textrm{q}} r^2}}{(1+\Z)^3}\l[r\HH_{_\textrm{q}}''(1+\Z)-r\HH_{_\textrm{q}}'\Z'+\HH_{_\textrm{q}}'(1+\Z)(4+3\Z)\r],\label{X21}\\
    \overset{2}{\Xbb}_2 &\circeq \frac{2}{3}\frac{r}{(1+\Z)^2}\l[-\dot\HH_{_\textrm{q}}'(1+\Z)+4(1+\Z)\HH_{_\textrm{q}}\HH_{_\textrm{q}}'+3r\HH_{_\textrm{q}}'^2 \r],\label{X22}\\
    \overset{2}{\Xbb}_3 &\circeq \frac{2}{5}\frac{\sqrt{1-\KK_{_\textrm{q}} r^2}}{(1+\Z)^3}\l[r\HH_{_\textrm{q}}''(1+\Z)-r\HH_{_\textrm{q}}'\Z'-\HH_{_\textrm{q}}'(1+\Z)(1+2\Z) \r]\label{X23},
\end{align}
As the LTB off-center observer is a specific case of the axisymmetric Szekeres off-center, we can immediately recover the LTB results of \cite{Sarma:2025yfw} for the multipoles of the covariant cosmographic parameters from \eqref{Hbb0} - \eqref{X23} by setting $\Z=0$, which is the only additional parameter that we have compared to the LTB set-up.\footnote{In \cite{Sarma:2025yfw}, we used the notation $H(t,r)=\HH_\textrm{q}(t,r)$ and $k(r)=\KK_{q0}(r)r^2$.} These multipoles depend on the location of the observer $r_o$. It must be emphasized that when $r_o=0$, all the derivatives of $\HH_\textrm{q}$ vanish owing to the regularity conditions at the center. Hence, a central observer finds only non-vanishing monopoles of the covariant cosmographic parameters, thereby indicating local isotropy. Also, the higher spatial derivative terms are associated to $\overset{2}{\Xbb}_1$ and $\overset{2}{\Xbb}_3$, which would increase the amplitudes of the dipole and octupole terms of the covariant deceleration \cite{Kalbouneh:2024yjj}. Lastly, we find that for FLRW background, all spatial derivatives vanish, we are left with monopoles and recover the standard cosmographic parameters of FLRW.

\subsection{Criteria for aligned multipoles}
Assuming $\HH_{_{q0}}'>0$ (which also makes $\Hbb_2>0$ for $|\Z|<1$) it is sufficient from (\ref{conditionsQ>0}) to have
\begin{align}\label{conditionsQ>0(2)}
    \frac{r\HH_{_{q0}}''}{\HH_{_{q0}}'}> \frac{1}{1+\Z}\l(r\Z'-(1+\Z)(4+3\Z)\r) \quad\textrm{and}\quad\frac{r\HH_{_{q0}}''}{\HH_{_{q0}}'}> \frac{1}{1+\Z}\l(r\Z'+(1+\Z)(1+2\Z)\r).
\end{align}

The alignment condition for the case of LTB reduces to 
\begin{align}\label{LTB-align}
    \frac{r\HH_{_{q0}}''}{\HH_{_{q0}}'}> 1 \quad (\textrm{for $\Qbb_1>0$ and $\Qbb_3>0$})\quad\textrm{or}\quad\frac{r\HH_{_{q0}}''}{\HH_{_{q0}}'}< -4 \quad (\textrm{for $\Qbb_1<0$ and $\Qbb_3<0$})
\end{align}

As the center of a Szekeres spacetime is a fixed point of isotropy, the kinematics of both the matter and null congruences can be locally characterized with the isotropic expansion of the congruences alone. Due to this, all the multipoles of the covariant cosmographic parameters except for the monopoles will vanish for a central observer.

\section{The Attractor-Repeller}\label{Attractor-Repeller}

The attractor–repeller is an aligned double-structure configuration consisting of an overdensity and an underdensity that appears to provide the best description of the peculiar velocity field, namely the deviations from a smooth FLRW background, as traced by the Cosmicflows-4 data \cite{Hoffman:2017ako}.

This configuration can naturally be analyzed within the framework of an axisymmetric $\Lambda$-Szekeres model. The line connecting the overdensity and the underdensity defines the symmetry axis, which we take to coincide with the (z)-axis. The observer is placed along this axis, between the two structures, so that the resulting covariant cosmographic parameters exhibit axisymmetric angular patterns. In the following subsections, we investigate a model consisting of a spherically symmetric overdensity and a pancake-like underdensity.

We are interested in an axisymmetric $\Lambda-$ Szekeres configuration that converges to a flat $\Lambda$CDM at spatial infinity.  As such, in the  modelling, we assume that $|A_{_\textrm{q}}'|>0$ for $r\neq0$ and $A_{_\textrm{q}}'=0$ only at $r=0$ and $r\to\infty$. Hence, the functional forms of the two remaining initial conditions, $\rho_{_\textrm{q0}}$ and $\Z$ can be written as
\begin{align}
    \rho_{_\textrm{q0}}(r)&= \bar\rho_{_0}\l(1+\delta_{_\textrm{q0}}(r)\r)\,,\label{rho_q0}\\
    \delta_{_\textrm{q0}}(r) &= 3 \delta_{_A} \left(\frac{R_{_A}}{r}\right)^3\left[\textrm{arcsinh}\left(\frac{r}{R_{_A}}\right)-\frac{1}{\sqrt{1+\left(\frac{R_{_A}}{r}\right)^2}}\right],\label{delta_q0}\\
    \Z(r) &= \delta_{_Z} \exp\left[-\frac{1}{2}\left(\frac{r-r_{_Z}}{R_{_Z}}\right)^2\right],
\end{align}
where $\bar\rho_{_0}$ is the $\Lambda$CDM matter density at $\tn$, $\delta_{_A}$ is the amplitude of density fluctuation of the spherically symmetric structure (structure-A) located at the center of the Szekeres with size $R_{_A}$. On the other hand, the pancake-like structure (structure-B) is located at the radial shell $r_{_Z}$ and characterized by an amplitude of density fluctuation $\delta_{_B}$ which is controlled by the amplitude of the dipole function $\delta_{_Z}$ and a size $\sigma_{_B}$ which is controlled indirectly by $R_{_Z}$.

We characterize the FLRW background using the parameters 
\begin{equation}
    \bar\Omega_{_\textrm{m0}} \equiv \lim_{r\to\infty}\Omega_{_\textrm{q0}}(r) =0.315, \quad\text{and}\quad  \Lambda =1.087\times 10^{-52}\; \text{m}^{-2}.
\end{equation}
These initial conditions on the background fix all the remaining variables of the background, \textit{viz.},
\begin{align}
     t_{_\textrm{0}}&=\frac{2}{\sqrt{3\Lambda}}\sinh^{-1}{\l(\sqrt{\frac{1-\bar\Omega_{_\textrm{m0}}}{\bar\Omega_{_\textrm{m0}}}}\r)} = 13.8\; \textrm{Gyr}\\
     \bar \rho_{_0} &\equiv \lim_{r\to\infty} \rho_{_{\textrm{q0}}}(r)= \frac{\Lambda}{8\pi G}\frac{\bar\Omega_{_\textrm{m0}}}{1-\bar\Omega_{_\textrm{m0}}} = 2.68\times 10^{-27}\; \textrm{kg/m}^3, \\
     \bar H_{_0} &\equiv \lim_{r\to\infty} \HH_{_{\textrm{q0}}}(r) = \sqrt{\frac{\Lambda }{3(1-\bar\Omega_{_\textrm{m0}})}} = 67.36\; \textrm{km/s/Mpc},\\
     \bar\KK_{_0}&\equiv \lim_{r\to\infty} \KK_{_{\textrm{q0}}}(r) =0,
\end{align}
which corresponds to the Planck values \cite{Planck_dip}
and they evolve as
\begin{align}
    \bar a(t) &\equiv \lim_{r\to\infty} a(t,r)= \l(\frac{\bar\Omega_{_\textrm{m0}}}{1-\bar\Omega_{_\textrm{m0}}}\r)^{1/3} \sinh^{2/3}{\l(\frac{3}{2}\sqrt{1-\bar\Omega_{_\textrm{m0}}}\; \bar H_{_0} t\r)},\\
    \bar H(t) &\equiv \lim_{r\to\infty} \HH_{_{\textrm{q}}}(t,r)= \frac{1}{\bar a(t)}\dv{\bar a(t)}{t} = \bar H_{_0} \sqrt{1-\bar\Omega_{_\textrm{m0}}} \coth{\l(\frac{3}{2}\sqrt{1-\bar\Omega_{_\textrm{m0}}}\; \bar H_{_0} t\r)},\label{Ht-FLRW}\\
    \bar\rho(t)&\equiv \lim_{r\to\infty} \rho_{_{\textrm{q}}}(t,r)= \bar\rho_{_0}\; \bar a^{-3}(t) = \frac{3\bar H_{_0}^2}{8\pi G}(1-\bar\Omega_{_\textrm{m0}})\sinh^{-2}{\l(\frac{3}{2}\sqrt{1-\bar\Omega_{_\textrm{m0}}}\; \bar H_{_0} t\r)},\label{rhot-FLRW}\\
    \bar\KK(t)&\equiv \lim_{r\to\infty} \KK_{_{\textrm{q}}}(t,r) =0\label{Kt-FLRW}.
\end{align}

\begin{figure}[htbp]
    \centering
    \includegraphics[scale=0.85]{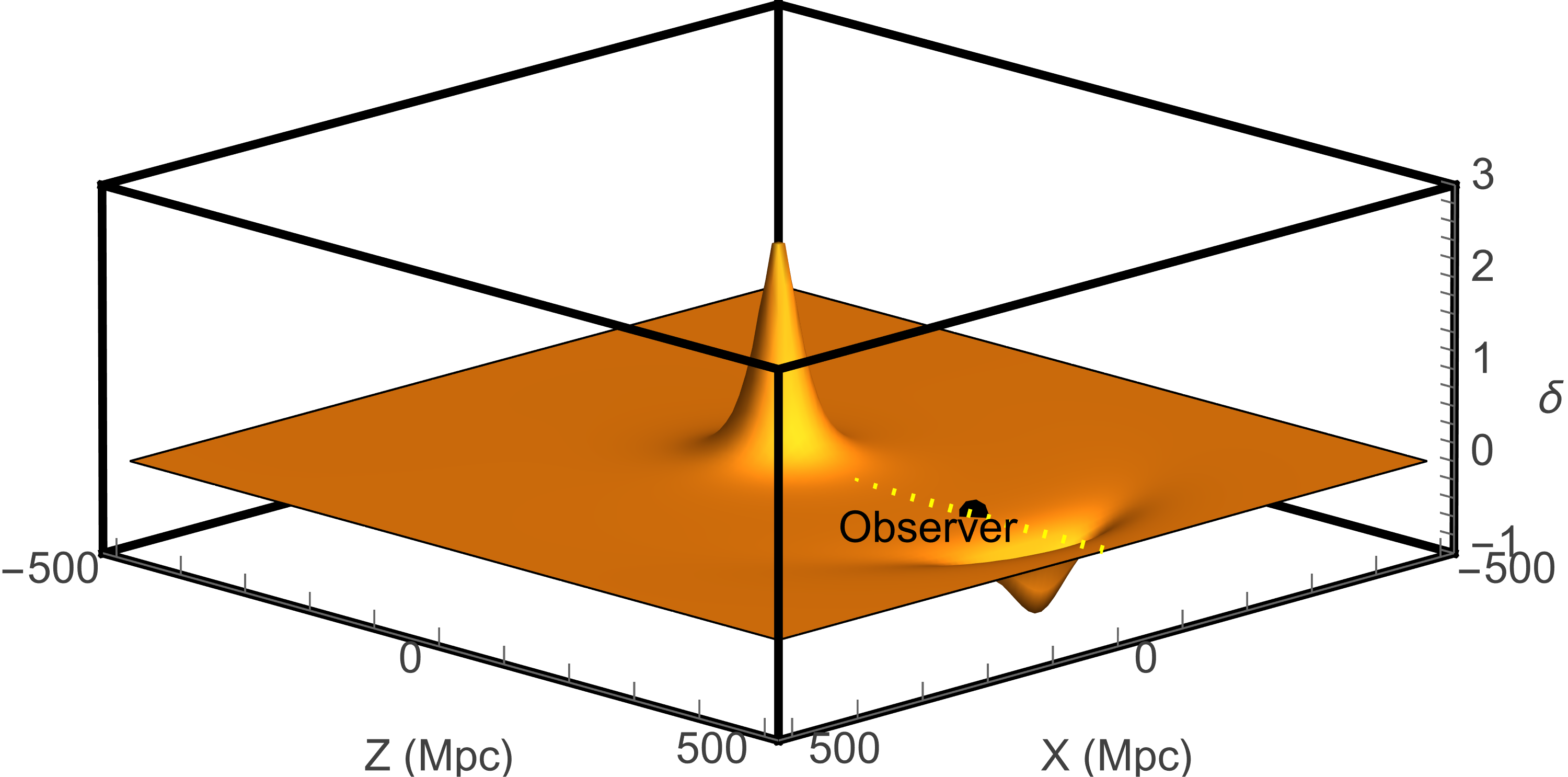}
    \caption{Density configuration of the attractor-repeller model with off-center observer.}
    \label{fig:density-1}
\end{figure}

Now that the background quantities at infinity are fixed, we characterise the local inhomogeneities.  
A spherically symmetric overdensity (attractor) of $\delta_{_A} = 2.5$ and size $R_{_A} = 37.4$ Mpc resembling the Shapley supercluster is placed at the center of the Szekeres configuration. A pancake like void (repeller) of $\delta_B \approx -0.5$ with size $\sigma_B \approx 40$ Mpc (FWHM) is placed along the $z$ axis $(\theta=0)$ centered at $r_{_Z}=400$ Mpc. This amplitude of the void is achieved with the dipole function parameters $\delta_{_Z}=-0.98$ and $R_{_Z} = 100$ Mpc. The density profile is shown in fig.(\ref{fig:density-1}).

\begin{figure}[htbp]
\centering
\hspace{-5mm}
\begin{subfigure}{0.5\textwidth}
    \centering
    \includegraphics[scale=0.47]{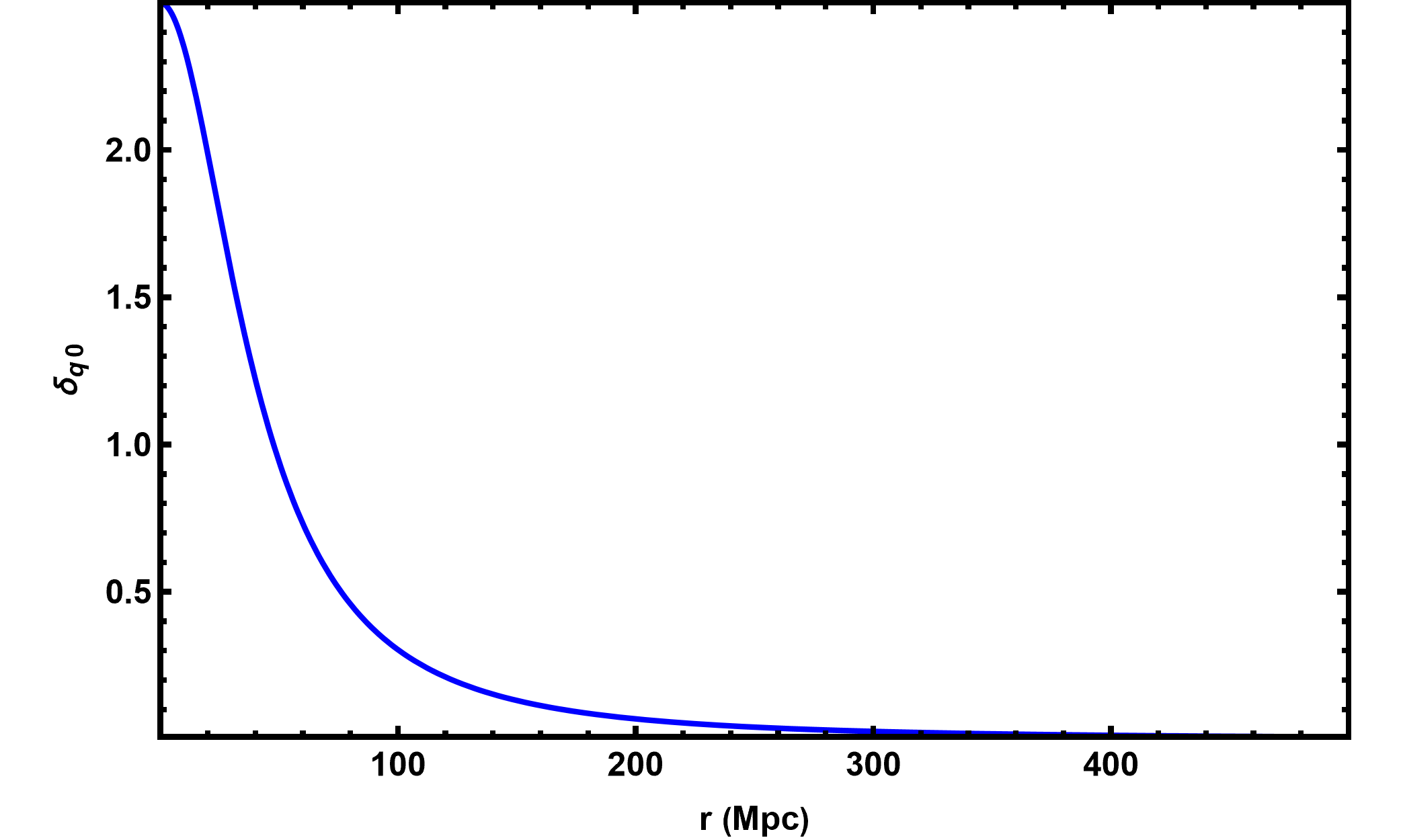}
    \subcaption{Profile of $\delta_{_\textrm{q0}}(r)$.}
\end{subfigure}
\hfill
\begin{subfigure}{0.5\textwidth}
    \centering
    \includegraphics[scale=0.47]{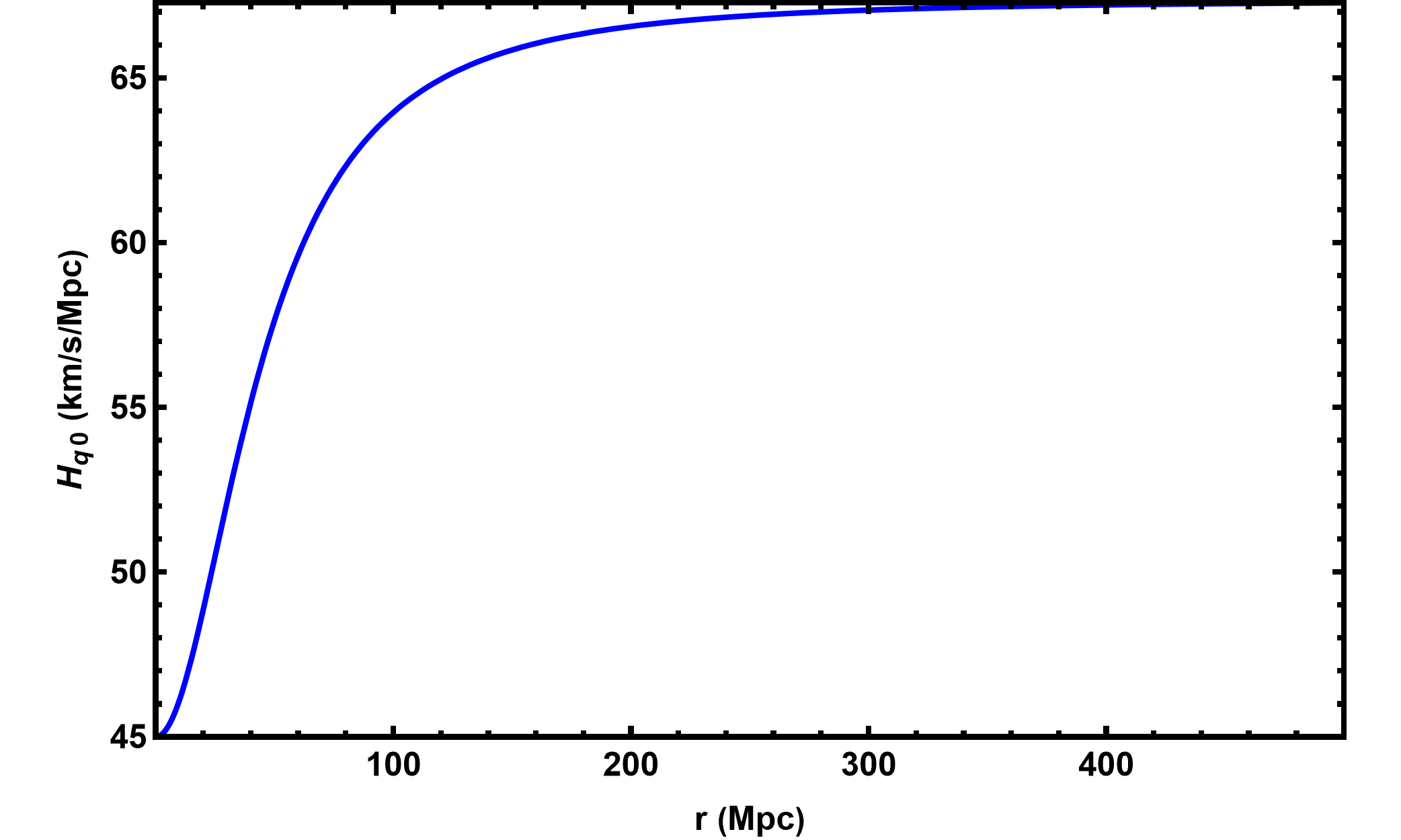}
    \subcaption{Profile of $\HH_{_\textrm{q0}}(r)$.}
\end{subfigure} 
\caption{Initial profiles of the q-scalars.}
\label{fig:q-1}
\end{figure}
We find that the non-vanishing dipole function near the void allows for the alignment condition (\ref{conditionsQ>0(2)}) to be satisfied. As such, an observer close to the void along the axis of symmetry at a radial coordinate of 300 Mpc satisfies these criteria for this specific profile. Increasing the size of the dipole function would increase the region of validity of the alignment of $\Qbb_1$ and $\Qbb_3$. However, near the center, this condition is solely determined by the background LTB model due to the vanishing dipole function. At the center, all the derivatives of the q-scalars must vanish, which would make all the multipoles except for the monopoles vanish. While changing the size of the dipole function, one must be careful to verify the regularity conditions at the center of the spacetime.

For the profile in (\ref{delta_q0}), we found that 
\begin{align}
    -4&<\frac{r\HH_{_{q0}}''}{\HH_{_{q0}}'}<1
\end{align}
irrespective of the sign of $\delta_A$. It is a very intriguing result, especially considering the non-linear regime. This is because, from \eqref{LTB-align}, we see that only for this range, the criterion for same sign of $\Qbb_1$ and $\Qbb_3$ is not satisfied. This alone rules out the specific single structured LTB profile in \cite{Sarma:2025yfw} regardless of the amplitude of density contrast. For the linear regime of LTB, it could be understood from the linearized Hubble profile (see \S 5 in \cite{Sarma:2025yfw})
\begin{align}
    \HH_{_{q0}}(r)&= \bar H_{_0}\l(1-\frac{1}{3}\delta_{_\textrm{q0}}(r) \r)\quad\implies\quad \frac{r\HH_{_{q0}}''}{\HH_{_{q0}}'}=\frac{r\delta_{_\textrm{q0}}''}{\delta_{_\textrm{q0}}'}
\end{align}
which cancels out the amplitude of density contrast from the numerator and denominator, making the expression insensitive to the sign of $\delta_A$. This result further strengthens the argument that a single structured LTB is incompatible with alignment of lower-order multipoles of the covariant cosmographic parameters. 

\begin{figure}[htbp]
    \centering
    \includegraphics[scale=0.55]{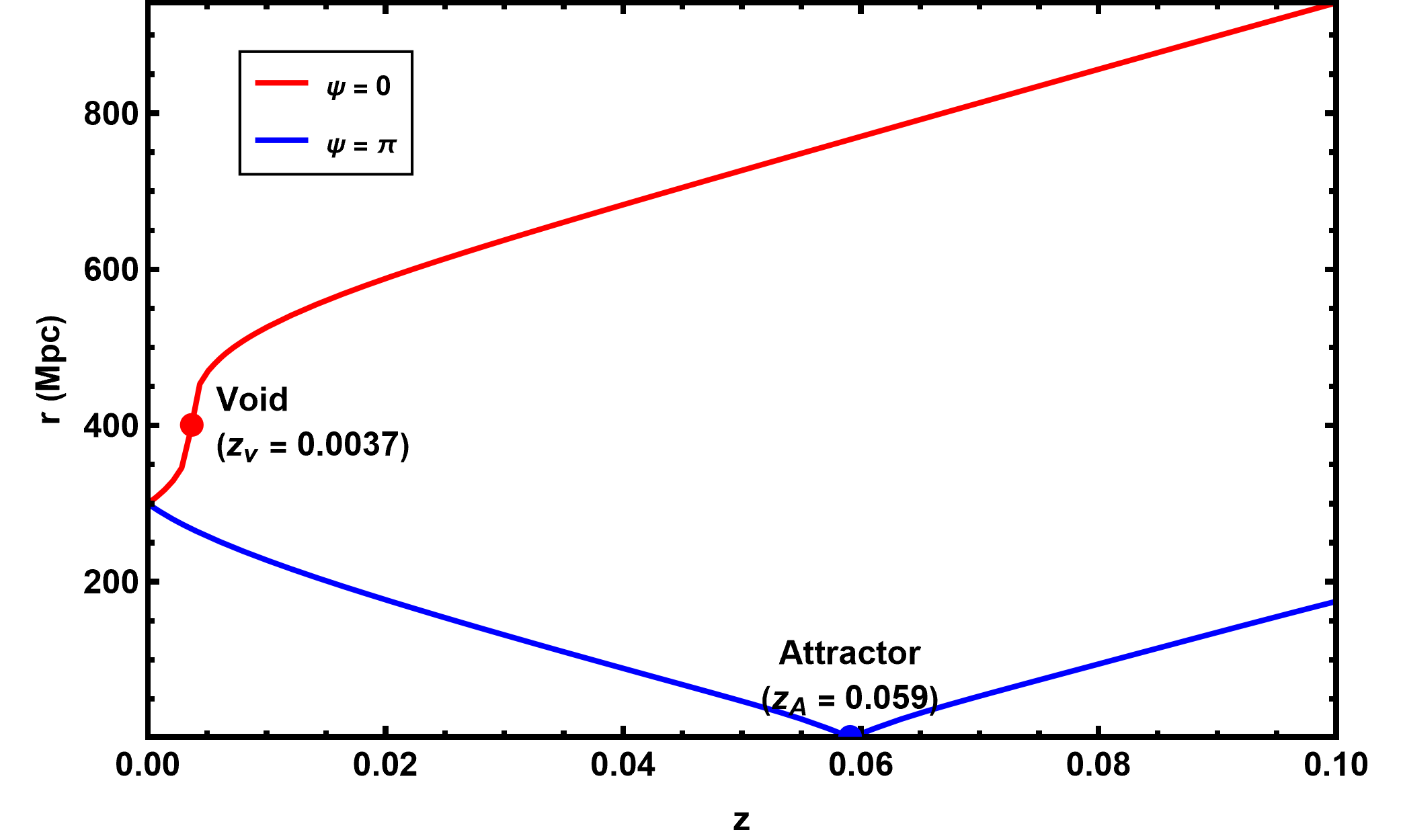}
    \caption{Radial coordinate $r$ as a function of redshift $z$ and line of sight $\psi$. $z_v$ is the redshift of the void and $z_A$ is the redshift of the attractor with respect to the off-center observer.}
    \label{fig:rcoord}
\end{figure}

As discussed by \cite{Sz-mult}, 
the radial coordinate can be very tricky to understand in the Szekeres spacetime. This is because, at any given time-slice, the foliations of the spacetime by 2-spheres of constant proper length along the radial direction is not concentric to the 2-spheres of constant $r$. As such, to understand physically the location of the structures labelled by $r$ coordinate shells, we relabel the $r$ coordinate in terms of redshift ($z$). To do so, we numerically integrate the null geodesics in the affine parameter $\lambda$. Moreover, as the redshift is specified from the position of the observer, so, the location of the two structures would be dependent on the line of sight as well with respect to the observer. For this case, we have the void along $\psi=0$ and the overdensity along $\psi=\pi$. The result is shown in fig.(\ref{fig:rcoord}).  

The dominant multipoles estimated by the observer upto $\mathcal{O}(z^2)$ are --
\begin{align}
    \Hbb_2/\Hbb_0 (10^{-2}) = 1.83\,,\quad
    \Qbb_1 &=2.01\,,\quad \Qbb_3=0.50\,.
\end{align}
It is clear from fig.(\ref{fig:q-1}), that in the attractor repeller double structure setup, the q-scalars vary monotonically in their first derivatives. As such, from (\ref{Hbb2}), the sign of the quadrupole is fixed regardless of the position of the observer. However, the magnitude of the derivatives, especially the higher order ones, would depend very much on the position of the observer. In this specific case, the sufficient conditions \eqref{conditionsQ>0(2)} for having the same sign of $\Qbb_1$ and $\Qbb_3$ is satisfied for an observer very close to the void ($d_L\approx 15.8$ Mpc).  But owing to the presence of a local structure very close by, it is expected that a covariant cosmographic expansion in the direction of the void would have a very small radius of convergence, $z<z_v \approx 0.004$. In fact, for this specific toy model, the higher-order spatial derivatives associated with the covariant jerk $\Jbb$ at $\mathcal{O}(z^3)$ would change rapidly with respect to the position of the observer due to the void nearby, making $\Jbb_2$ and  $\Jbb_4$ unstable. Hence, $\Jbb$ is not discussed further in this model. Furthermore, we note that there is a difference of almost 2.5 $\sigma$  in the calculated value of $\Hbb_2/\Hbb_0$ with respect to the CF4 value. However, we do not consider this discrepancy significant, as our aim is to qualitatively explore the configurations compatible with the same sign of the covariant deceleration multipoles, rather than to perform a fit.
\begin{figure}[htbp]
    \centering
    \includegraphics[scale=0.55]{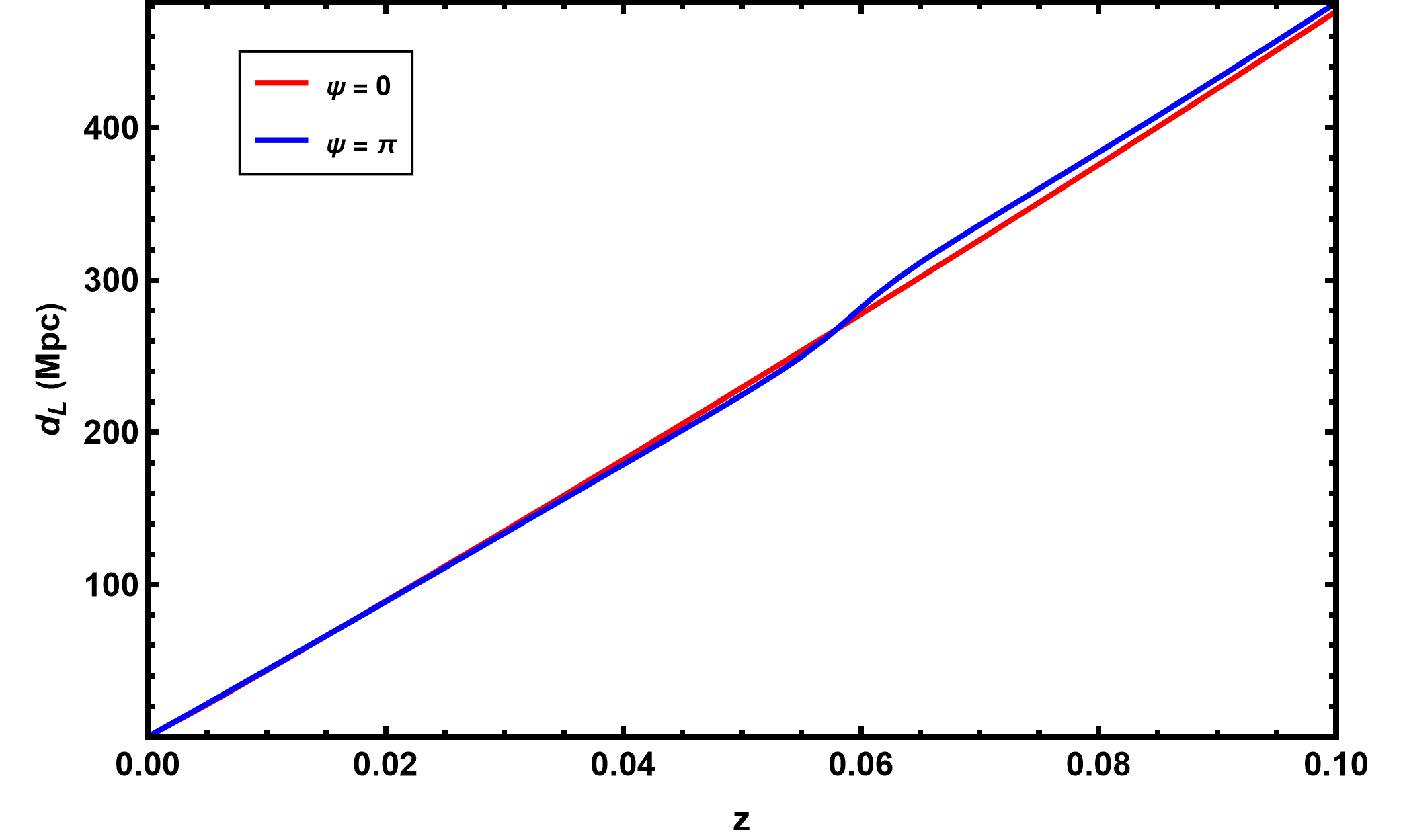}
    \caption{Luminosity distance for the off-center observer in the model-1. The scaling with redshift along two different line-of-sight directions is shown: towards the void ($\psi=0$, solid red line) and towards the overdensity ($\psi=\pi$, solid blue line).}
    \label{fig:dlexact-1}
\end{figure}

\section{Relation to Cosmological Perturbation Theory}\label{sec-Perturbation Theory}

We identify the Szekeres spacetime with its asymptotic FLRW background through the same comoving synchronous threading, defining a common background identification of the constant-t hypersurfaces
\begin{equation}
    \Sigma_t=\{p\in\mathcal{M} : t(p)=\text{constant}\}
\end{equation}
with the induced 3-metrics $h_{ab}=g_{ab} + u_a u_b$ (Szekeres) and  $\bar h_{ab}=\bar g_{ab} + u_a u_b$ (FLRW) defined on that identical $\Sigma_t$. 

The consequence of this is that the covariant scalars $(\rho,\HH,\KK)$ can be split into an exact fluctuation about their respective background FLRW values $(\bar\rho,\bar\HH,\bar\KK)$ in the same $t = $ constant slice (and not across different slices of $t$) as given in \eqref{Ht-FLRW}-\eqref{Kt-FLRW}. Because the Szekeres spacetime has pressureless dust as a source along with $\Lambda$, the corresponding background FLRW is a flat $\Lambda$CDM. 

We can consider the Szekeres inhomogeneities in the language of linear scalar perturbations about the flat $\Lambda$CDM background. A detailed analysis was presented in \cite{Sz-CPT}, where the authors established a correspondence with scalar perturbations of the FLRW spacetime in the comoving synchronous gauge. Our goal, instead, is to establish the corresponding connection with the variables of the Newtonian gauge. To this end, we explicitly compute the gauge-invariant gravitational potential $\Phi$ and the peculiar velocity $\mathbf{v}_{\mathrm{p}}$ from the density inhomogeneity of the axisymmetric Szekeres structure.

The  density fluctuation about the asymptotic FLRW  is  
\begin{equation}\label{eq-density contr}
    \delta(t,{\bf r}) \equiv \frac{\rho(t,{\bf r})-\bar\rho(t)}{\bar\rho(t)}
\end{equation}
with $\rho$ the density of the Szekeres spacetime and $\bar\rho$  its asymptotic FLRW value at the constant time hypersurface $\Sigma_t$. It is evident from the definitions  \eqref{A} and \eqref{DDA}, that the q-scalars should also correspond to the same background FLRW values for the same $\Sigma_t$ (owing to the vanishing of the spatial derivatives of the q-scalars at spatial infinity). Thus,
\begin{equation}\label{eq- q density contr}
    \rho_{_\textrm{q}}(t,r) = \bar\rho(t)\Big( 1+\delta_{\textrm{q}}(t,r)\Big) \quad \implies \quad \delta_\textrm{q}(t,r) \equiv \frac{\rho_\textrm{q}(t, r)-\bar\rho(t)}{\bar\rho(t)}
\end{equation}
Together with \eqref{A},  \eqref{eq-density contr} and \eqref{eq- q density contr}
\begin{equation}
    \delta(t,{\bf r}) = \delta_{_\textrm{q}}(t,r) + \frac{r \delta_{_\textrm{q}}'(t,r)}{3\Big(\Gamma(t,r)-\Wb(r,\theta,\phi) \Big)}.
\end{equation}
In linear perturbation theory, the exact density fluctuation $\delta$ would give rise to the gauge invariant gravitational potential $\Phi(t,{\bf r})$ obtained by integrating the Poisson equation and a peculiar velocity field ${\bf v_p}(t,{\bf r})$ which is the gradient of $\Phi(t,{\bf r})$
\begin{align}
    \Phi(t,{\bf r}') &= \N \int \dd \V\; \delta(t,{\bf r}) \frac{1}{|\textbf{r}' - \textbf{r}|},\quad\text{where}\quad \N= -\frac{3\bar\Omega_{_\textrm{m}}\bar H^2\bar a^2}{8\pi},\label{eq-Phi}\\
    {\bf v_p}(t,{\bf r}') &= \M\;\boldsymbol{\nabla}\Phi(t,{\bf r}')\quad\text{where}\quad \M=-\frac{2 f(\bar\Omega_{\textrm{m}})}{3\bar H\bar\Omega_\textrm{m}\bar a}\label{eq-vp},
\end{align}
and $f(\bar\Omega_{_\textrm{m}})=\bar\Omega_{_\textrm{m}}^{0.55}$ is the growth rate \footnote{$\boldsymbol{\nabla}$ is the spatial derivative.}. Note that this assumes that the exact structure is in sub-horizon scales which is true for our model.

\begin{figure}[htpb]
    \centering
    \includegraphics[scale=1.1]{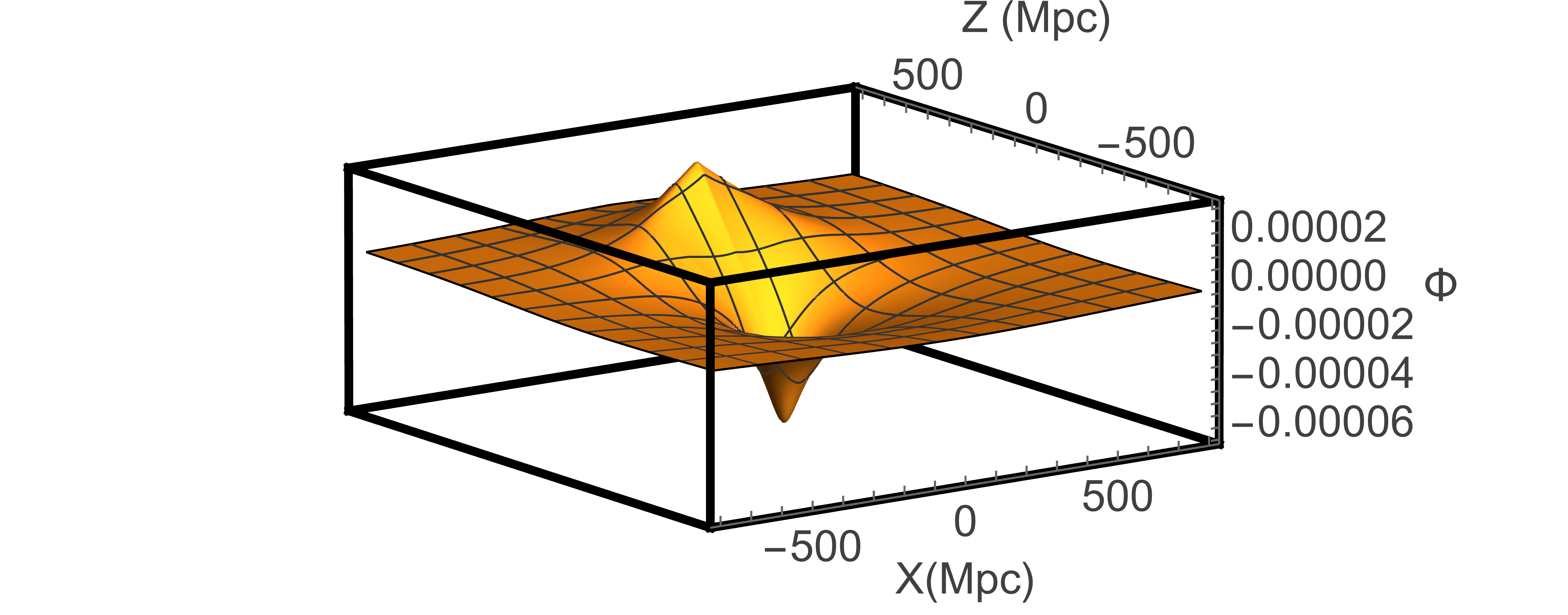}
    \caption{Gravitational potential $\Phi$ associated to the density $\delta_{_0}$ in Szekeres.}
    \label{fig:Phi}
\end{figure}

Since we are interested only in the present-day values of $\Phi$ and ${\bf v_p}$  corresponding to the density profile in \S \ref{Attractor-Repeller}, \textit{i.e.},
\begin{equation}\label{eq-delta_0}
    \delta_{_0}(r,\theta) = \delta_{_\textrm{q0}}(r) + \frac{r\delta_{_\textrm{q0}}'(r)}{3\Big(1+\Z(r) \cos{\theta}\Big)}.
\end{equation}
we omit the subscript "$_0$" from here onward for brevity, with the understanding that $\Phi$ and ${\bf v_p}$ are evaluated at the present epoch $t_{_0}$. Hence, using \eqref{eq-delta_0} in \eqref{eq-Phi}, we get
\begin{align}
    \Phi(r',\theta') &= \mathcal{N}\int \dd \textbf{r}\; \delta_0(r,\theta) \frac{1}{|\textbf{r}' - \textbf{r}|}\\
    &=\N\int \dd \textbf{r}\; \left[\delta_{_\textrm{q0}}+\frac{r\delta_{_\textrm{q0}}'}{3}\Big(1+\Z\cos{\theta}\Big)^{-1}\right]\frac{1}{|\textbf{r}' - \textbf{r}|}
\end{align}
Because in our models, we avoid shell crossing by setting $|\Z|<1$, as such, one can do a binomial expansion in $\Z \cos{\theta}$ for the density profile, \textit{i.e.},
\begin{equation}
   \Big( 1+\Z\cos{\theta}\Big)^{-1}\approx 1-\Z\cos{\theta}
\end{equation}
to see the leading order correction due to Szekeres dipole function about an LTB model. It results in
\begin{align}
    \Phi(r',\theta') &\approx \N \int \dd \textbf{r}\; \left[\delta_{_\textrm{q0}}+\frac{r\delta_{_\textrm{q0}}'}{3}-\frac{r\delta_{_\textrm{q0}}'}{3}\Z\cos{\theta}\right]\frac{1}{|\textbf{r}' - \textbf{r}|}\\
    &\equiv \Phi_{\textrm{LTB}}  + \Phi_{\textrm{Sz-corr}}
\end{align}
where 
\begin{align}
     \Phi_{\textrm{LTB}}(r')&\equiv \N \int \dd \textbf{r}\; \left[\delta_{_\textrm{q0}}+\frac{r\delta_{_\textrm{q0}}'}{3}\right]\frac{1}{|\textbf{r}' - \textbf{r}|} \label{Phi-ltb}\\
     \Phi_{\textrm{Sz-corr}}(r',\theta')&\equiv \N \int \dd \textbf{r}\; \left[-\frac{r\delta_{_\textrm{q0}}'}{3}\Z \cos{\theta}\right]\frac{1}{|\textbf{r}' - \textbf{r}|}\label{Phi-Sz}
\end{align}
Using 
\begin{equation}
    \frac{1}{|\textbf{r}' - \textbf{r}|}= \sum_{l=0}^\infty \frac{r_{<}^l}{r_{>}^{l+1}} P_l(\cos{\gamma}) 
\end{equation}
where $r_<\equiv \textrm{min}\{{r,r'}\}$, $r_>\equiv \textrm{max}\{{r,r'}\}$ and  $\cos{\gamma}\equiv \cos\theta\cos\theta' + \sin\theta\sin\theta' \cos{(\phi-\phi')}$ and 
\begin{equation}
     P_l(\cos{\gamma}) = \frac{4\pi}{2l+1}\sum_{m=-l}^l Y_{lm}(\theta',\phi')Y^*_{lm}(\theta,\phi)
\end{equation}
we can write \eqref{Phi-ltb} as
\begin{align}
    \Phi_{\textrm{LTB}}(r')&= \N \int \dd \V \left[\delta_{_\textrm{q0}}+\frac{r\delta_{_\textrm{q0}}'}{3}\right] \sum_{l=0}^\infty\sum_{m=-l}^l\Bigg(\frac{4\pi}{2l+1}\Bigg(\frac{r_<^l}{r_>^{l+1}}\Bigg)Y_{lm}(\theta',\phi')Y^*_{lm}(\theta,\phi)\Bigg)\\
    &= \N \int \dd r\; r^2 \left[\delta_{_\textrm{q0}}+\frac{r\delta_{_\textrm{q0}}'}{3}\right] \sum_{l=0}^\infty\sum_{m=-l}^l\Bigg(\frac{4\pi}{2l+1}\Bigg(\frac{r_<^l}{r_>^{l+1}}\Bigg)Y_{lm}(\theta',\phi')\int \dd\Omega\; Y^*_{lm}(\theta,\phi)\Bigg)
\end{align}
Using $\int \dd\Omega\; Y^*_{lm}(\theta,\phi) = \sqrt{4\pi}\;\delta_{l0}\delta_{m0}$ and substituting the delta functions and $Y_{00}=1/\sqrt{4\pi}$, the above integral simplifies to
\begin{align}
    \Phi_{\textrm{LTB}}(r') &= 4\pi\N\int_0^\infty \dd r\; \frac{r^2}{r_>}\Bigg[\delta_{_\textrm{q0}}+\frac{r\delta_{_\textrm{q0}}'}{3}\Bigg]\\
    &= 4\pi\N\int_0^{r'} \dd r\; \frac{r^2}{r'}\Bigg[\delta_{_\textrm{q0}}+\frac{r\delta_{_\textrm{q0}}'}{3}\Bigg] + 4\pi\N\int_{r'}^\infty \dd r\; \frac{r^2}{r}\Bigg[\delta_{_\textrm{q0}}+\frac{r\delta_{_\textrm{q0}}'}{3}\Bigg]\\
    &= \frac{4\pi}{3r'}\N\int_0^{r'}\dd\Big(r^3\delta_{\textrm{q0}}\Big)+4\pi\N\int_{r'}^\infty \dd r\; r\Bigg[\delta_{_\textrm{q0}}+\frac{r\delta_{_\textrm{q0}}'}{3}\Bigg]\\
    &= \frac{4\pi}{3r'}\N\int_0^{r'}\dd\Big(r^3\delta_{\textrm{q0}}\Big) + \frac{4\pi}{3}\N\int_{r'}^\infty\dd\Big(r^2\delta_{\textrm{q0}}\Big) + \frac{4\pi}{3}\N\int_{r'}^\infty\dd r\; r \delta_{\textrm{q0}}
\end{align}
The first term can be cast as a total derivative and since $\displaystyle\lim_{r\to\infty}\Big(r^2\delta_\textrm{q0}\Big)=0$, the first two terms exactly cancel out and we are left with
\begin{equation}\label{phi-LTB}
    \Phi_{\textrm{LTB}}(r') = \frac{4\pi}{3}\N\int_{r'}^\infty\dd r\; r\, \delta_{\textrm{q0}}(r) = -\frac{1}{2}\bar\Omega_\textrm{m0}\bar H_0^2\;\int_{r'}^\infty\dd r\; r\, \delta_{\textrm{q0}}(r).
\end{equation}
It is interesting to see that the gravitational potential constructed out of an LTB density contrast $\delta_0$ would be affected only by its quasi local density contrast $\delta_\textrm{q0}$. Moreover, if we calculate the peculiar velocity field associated to $\Phi_\textrm{LTB}$ at $t_0$, we find that
\begin{equation}\label{vp-ltb}
    {\bf v_p}_{\,\textrm{LTB}} =\M\;\boldsymbol{\nabla}\Phi_\textrm{LTB}(r) =  -\frac{1}{3}f\bar H_0 r\delta_\textrm{q0}\, \hat {\bf r}
\end{equation}
which is the classical infall formula of cosmological perturbation theory. 
Now, going back to the non-spherical Szekeres correction term in the potential  \eqref{Phi-Sz}, we have,

\begin{align}
    \Phi_{\textrm{Sz-corr}}(r',\theta')&= \N \int \dd \V \left[-\frac{r\delta_{_\textrm{q0}}'}{3}\Z\cos{\theta}\right] \sum_{l=0}^\infty\sum_{m=-l}^l\Bigg(\frac{4\pi}{2l+1}\Bigg(\frac{r_<^l}{r_>^{l+1}}\Bigg)Y_{lm}(\theta',\phi')Y^*_{lm}(\theta,\phi)\Bigg)\\
    &= \N \int \dd r\; r^2 \left[-\frac{r\delta_{_\textrm{q0}}'}{3}\Z\right] \sum_{l=0}^\infty\sum_{m=-l}^l\Bigg(\frac{4\pi}{2l+1}\Bigg(\frac{r_<^l}{r_>^{l+1}}\Bigg)Y_{lm}(\theta',\phi')\int \dd\Omega\;\Big(\cos{\theta}\, Y^*_{lm}(\theta,\phi)\Big)\Bigg)
\end{align}
Using $\int \dd\Omega\; Y^*_{lm}(\theta,\phi)\cos{\theta} = \sqrt{4\pi/3}\;\delta_{l1}\delta_{m0}$ and substituting the delta functions and $Y_{10}(\theta',\phi')=\sqrt{3/(4\pi)}\cos{\theta'}$, the above integral simplifies to
\begin{align}
    \Phi_\textrm{Sz-corr}(r',\theta')&= \frac{4\pi}{3}\N\cos{\theta'}\int_0^\infty \dd r\;\Bigg(-\frac{r^3\delta_{\textrm{q0}}'}{3}\Z\Bigg)\frac{r_<}{r_>^2}\\
    &= \frac{4\pi}{3}\N\cos{\theta'}\Bigg[\int_0^{r'} \dd r\;\Bigg(-\frac{r^3\delta_{\textrm{q0}}'}{3}\Z\Bigg)\frac{r}{r'^2} + \int_{r'}^\infty \dd r\;\Bigg(-\frac{r^3\delta_{\textrm{q0}}'}{3}\Z\Bigg)\frac{r'}{r^2}\Bigg]\\
     &=-\frac{1}{2}\bar\Omega_\textrm{m0}\bar H_0^2\;\cos{\theta'}\Bigg[\int_0^{r'} \dd r\;\Bigg(-\frac{r^3\delta_{\textrm{q0}}'}{3}\Z\Bigg)\frac{r}{r'^2} + \int_{r'}^\infty \dd r\;\Bigg(-\frac{r^3\delta_{\textrm{q0}}'}{3}\Z\Bigg)\frac{r'}{r^2}\Bigg]
\end{align}
The peculiar velocity associated to $\Phi_\textrm{Sz-corr}$ is 
\begin{align}
    {\bf v_p}_{\,\textrm{Sz-corr}}(r',\theta') &= \M\;\boldsymbol{\nabla}\Phi_\textrm{Sz-corr}(r,\theta)\\
    &= -\frac{1}{3}f(\bar\Omega_\textrm{m0}) \frac{\sin{\theta'}}{r'}\Bigg[\int_0^{r'} \dd r\;\Bigg(-\frac{r^3\delta_{\textrm{q0}}'}{3}\Z\Bigg)\frac{r}{r'^2} + \int_{r'}^\infty \dd r\;\Bigg(-\frac{r^3\delta_{\textrm{q0}}'}{3}\Z\Bigg)\frac{r'}{r^2}\Bigg] \hat{\boldsymbol{\theta}}
\end{align}
Hence, the peculiar velocity field in an axisymmetric Szekeres structure can be decomposed into a radial component owing to the LTB structure and angular component owing to the Szekeres correction. For the specific attractor-repeller model in \S\ref{Attractor-Repeller}, the gravitational potential $\Phi$ at $t_{_\textrm{0}}$ is shown in fig.\eqref{fig:Phi}. The shape of this profile is as expected from linear perturbation theory. 

\begin{figure}[htpb]
    \centering
    \includegraphics[scale=0.5]{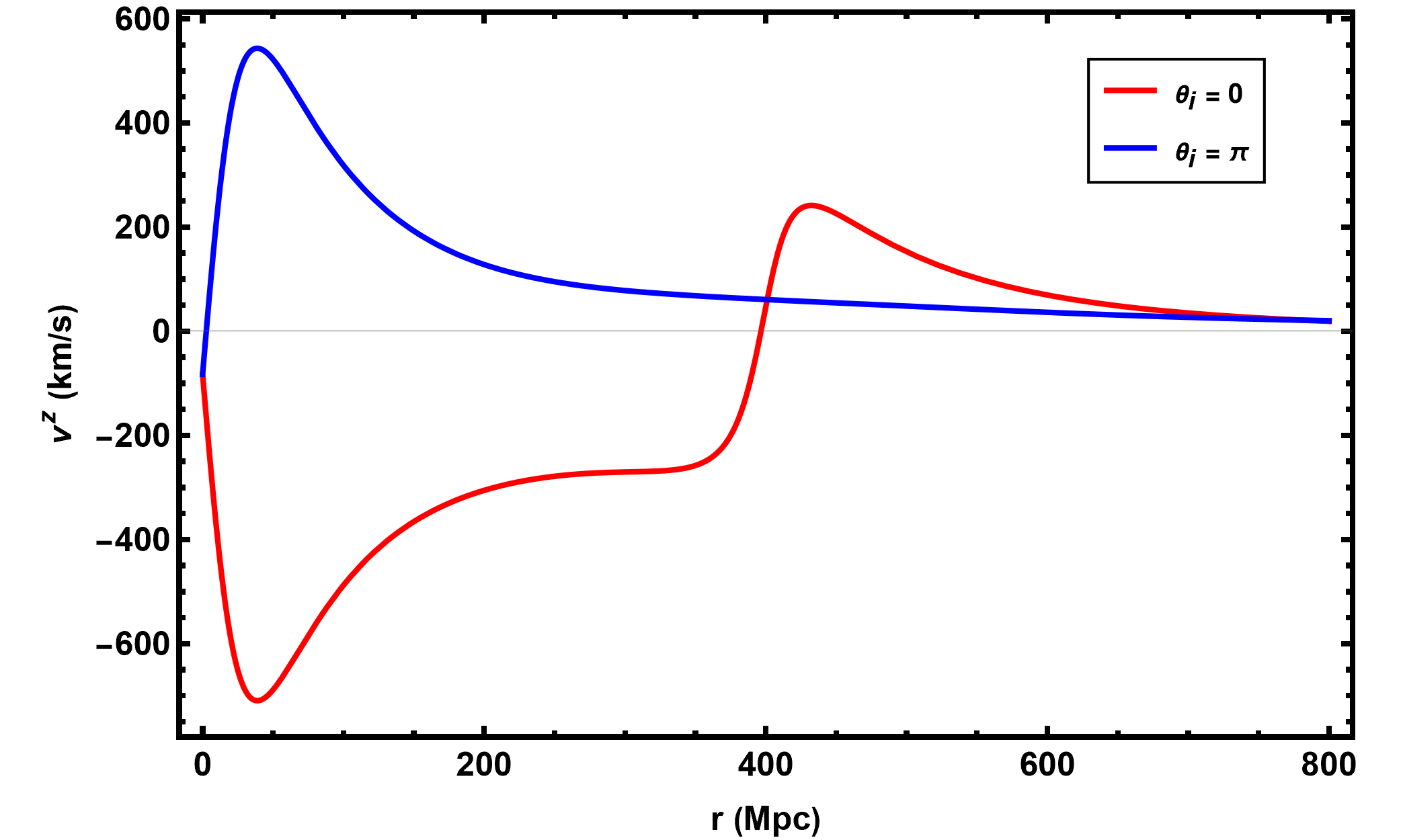}
    \caption{Peculiar velocity $\textbf{v} (=v^z\; \hat z)$ along the axis of symmetry.}
    \label{fig:vp}
\end{figure}

The peculiar velocity profile along the $z$ axis is shown in fig.\eqref{fig:vp}. We notice from this plot that at the position of the observer $r_o=300$ Mpc, the peculiar velocity is about 285 km/s (the plot in red). The negative sign suggests that the direction of the velocity is towards the center which is expected from this attractor-repeller set-up \footnote{Along the +ve z axis ($\theta_i=0$), $v^z>0$ corresponds to a velocity directed radially away from the attractor (located at the center) and $v^z<0$ corresponds to a velocity directed towards it. Along the -ve z axis ($\theta_i=\pi$), the situation is reversed.}. At the center of the structure, $v^z$ is about 100 km/s along the -ve z axis. Along $\theta_i=0$, $v^z$ changes the sign at 400 Mpc owing to the void.   

The full Szekeres correction in $\Phi$ and ${\bf v_p}$, owing to the expansion
\begin{equation}
    \Big(1+\Z\cos{\theta}\Big)^{-1} = \sum_{s=0}^\infty (-1)^s(\Z\cos{\theta})^s
\end{equation}
can be written as 
\begin{align}
    \Phi_\textrm{Sz-corr-full}(r',\theta') &= \N \sum_{s=1}^\infty (-1)^s \sum_{l=0}^\infty\sum_{m=-l}^l\Bigg[\frac{4\pi}{2l+1}Y_{lm}(\theta',\phi')\int \dd\Omega\;\Big(\cos^s\theta\, Y^*_{lm}(\theta,\phi)\Big)\int \dd r\;\left[\frac{r^3\delta_{_\textrm{q0}}'}{3}\Z^s\right] \Bigg(\frac{r_<^l}{r_>^{l+1}}\Bigg)\Bigg]
\end{align}
Using further $Y_{l0}=\sqrt{(2l+1)/(4\pi)}\;P_l(\cos\theta)$ and 
\begin{equation}
    \int \dd\Omega\;\Big(\cos^s\theta\, Y^*_{lm}(\theta,\phi)\Big)= 2\pi \sqrt{\frac{2l+1}{4\pi}}\int_{-1}^1\dd y\; y^s P_l(y)\quad \text{where}\quad y=\cos\theta.
\end{equation}
we get
\begin{align}
    \Phi_\textrm{Sz-corr-full}(r',\theta') &= 2\pi \N \sum_{s=1}^\infty (-1)^s \sum_{l=0}^\infty\Bigg[P_l(\cos{\theta'})\int_{-1}^1 \dd y\;y^sP_l(y)\int \dd r\;\left[\frac{r^3\delta_{_\textrm{q0}}'}{3}\Z^s\right] \Bigg(\frac{r_<^l}{r_>^{l+1}}\Bigg)\Bigg].
\end{align}
This give us
\begin{align}
    {\bf v_p}_{\,\textrm{Sz-corr-full}}(r',\theta') &= \mathcal{M}\, \boldsymbol{\nabla}\Phi_\textrm{Sz-corr-full}(r',\theta')\\
    &= \mathcal{M}\;\frac{1}{r'}\pdv{\Phi_\textrm{Sz-corr-full}}{\theta'}\;\hat{\boldsymbol{\theta}}\\
    &=\frac{2\pi\mathcal{M}\N}{r'} \sum_{s=1}^\infty (-1)^s \sum_{l=0}^\infty\Bigg[\dv{P_l(\cos{\theta'})}{\theta'}\int_{-1}^1 \dd y\;y^sP_l(y)\int \dd r\;\left[\frac{r^3\delta_{_\textrm{q0}}'}{3}\Z^s\right] \Bigg(\frac{r_<^l}{r_>^{l+1}}\Bigg)\Bigg] \hat{\boldsymbol{\theta}}\\
    &= \frac{2\pi\mathcal{M}\N}{r'} \sum_{s=1}^\infty (-1)^s \sum_{l=0}^\infty\Bigg[\frac{l+1}{\sin{\theta'}}\Big(\cos{\theta}P_l(\cos\theta)-P_{l+1}(\cos\theta)\Big)\int_{-1}^1 \dd y\;y^sP_l(y)\nonumber\\
    &\quad \int \dd r\;\left[\frac{r^3\delta_{_\textrm{q0}}'}{3}\Z^s\right] \Bigg(\frac{r_<^l}{r_>^{l+1}}\Bigg)\Bigg]\;\hat{\boldsymbol{\theta}}
\end{align}
\subsection{Consistency check: Linear LTB as Perturbed FLRW with EdS background}
A detailed study of linearized LTB model as linearly perturbed FLRW with an EdS background has been carried out in \cite{Sarma:2025yfw}. Here, we propose that instead, if one starts with \eqref{eq-Phi} and \eqref{eq-vp} for a linearized LTB solution that converges to an EdS background at spatial infinity, then one recovers the same relations as \cite{Sarma:2025yfw} between the parameters of the linearized LTB solution and the parameters of the FLRW with scalar perturbations about the EdS background. 

The Class I quasi-spherical axisymmetric Szekeres models reduce to LTB if $\Z(r)=0$. Moreover, in a linearized LTB, we have
\begin{align}
    \HH_{_\textrm{q0}}(r) &\approx \bar H_0\Bigg(1-\frac{1}{3}\delta_{_\textrm{q0}}(r)\Bigg) \quad \text{and} \quad \Omega_{_\textrm{q0}}(r)\approx 1 + \frac{5}{3}\delta_{_\textrm{q0}}(r)
\end{align}
Using the linearized Hamiltonian constraint, 
\begin{align}\label{eq-EdS K}
    \KK_{_\textrm{q0}}(r) &\approx \frac{5}{3}\bar H_0^2 \delta_{_\textrm{q0}}(r).
\end{align}
From \eqref{vp-ltb}, we get the relation between the curvature $\KK_\textrm{q0}$ and the radial component of peculiar velocity $v_p^r = - \bar H_0 r \delta_\textrm{q0}/3$ \footnote{$f(\bar\Omega_{m0})=1$ for EdS background.} as
\begin{equation}
    \KK_{_\textrm{q0}} =-\frac{5\bar H_0^2}{r}v_p^r,
\end{equation}
which is the exact relation in (5.49) of \cite{Sarma:2025yfw}. Hence, we verify the consistency of our results.

\section{Conclusion}
In this work, we implemented the framework of a quasi-spherical axisymmetric Szekeres I spacetime with dust and $\Lambda$ to explain the observed axisymmetric expansion rate. We applied the covariant cosmographic (CC) framework by placing an off-center observer on the axis of symmetry and estimated the multipoles of the CC parameters up to the deceleration, $\mathcal{O}(z^2)$. We derived the sufficient conditions for having a positive quadrupole of the covariant Hubble parameter, along with the dipole and octupole of covariant deceleration having the same sign. Further, we calculated the luminosity distance for the off-center observer by integrating the Sachs equations. We emphasize that these results could be implemented in multiple networks of axisymmetric structures in quasi-spherical Szekeres class I as well as multiple structures in LTB spacetimes by setting the dipole function to zero.   

The axisymmetric Szekeres structure has an intrinsic mass dipole which gives rise to anisotropic gravitational and velocity fields. Moreover, the formalism correctly reduces to the LTB limit when the Szekeres dipole function vanishes.

As an application of this framework, we investigated an attractor-repeller system constructed by a spherically symmetric overdensity resembling the Shapley at the center of the spacetime and a pancake-like void. The observer is placed between the two structures. We estimated the amplitude of the covariant cosmographic parameters numerically. Due to the presence of a very nearby structure (the void), the radius of convergence of the cosmographic luminosity distance is expected to be very small. 

Following this, we examined the exact luminosity distance--redshift relation along the two preferred directions defined by the local structures, namely toward the void and toward the overdensity. To further relate the formalism to Cosmological Perturbation Theory, we derived the expressions for the gravitational potential and the peculiar velocity associated to the exact density fluctuation in the Szekeres spacetime. The peculiar velocity of the observer was found to be 285 km/s. We note that this is a genuine prediction of the setup, as we have no further freedom in determining the velocity profile once the Szekeres initial data are fixed. Interestingly, even in this exact inhomogeneity, we found a small $\Phi$ and ${\bf v_p}$. This explains why perturbative observables remain applicable in such setups.

At this stage, the specific limitations of the model are the factor of 1.5 discrepancy in the calculated value of $\Hbb_2/\Hbb_0$ with the CF4 value and the small radius of convergence near the void. Nevertheless, the qualitative agreement obtained with a simple parametrization is encouraging. The natural follow-up to this work is a proper MCMC fit to CF4 data, with the four free parameters of the attractor-repeller setup left to vary, to determine their best-fit values.

\acknowledgments 

We are grateful to Roberto A. Sussman for insightful comments on interpreting peculiar velocities in Szekeres solutions and engaging discussions. We would also like to thank Marie-Noëlle Célérier, Jessica Santiago, Roy Maartens and Chris Clarkson for useful discussions.  MS, CM and BK are supported by the {\it Agence Nationale de la Recherche} under the grant ANR-24-CE31-6963-01, and the French government under the France 2030 investment plan, as part of the Initiative d’Excellence d'Aix-Marseille Université -  A*MIDEX (AMX-19-IET-012).

\appendix
\section{Light propagation and Sachs equations}
\begin{figure}[H]
    \centering
    \includegraphics[scale=0.1]{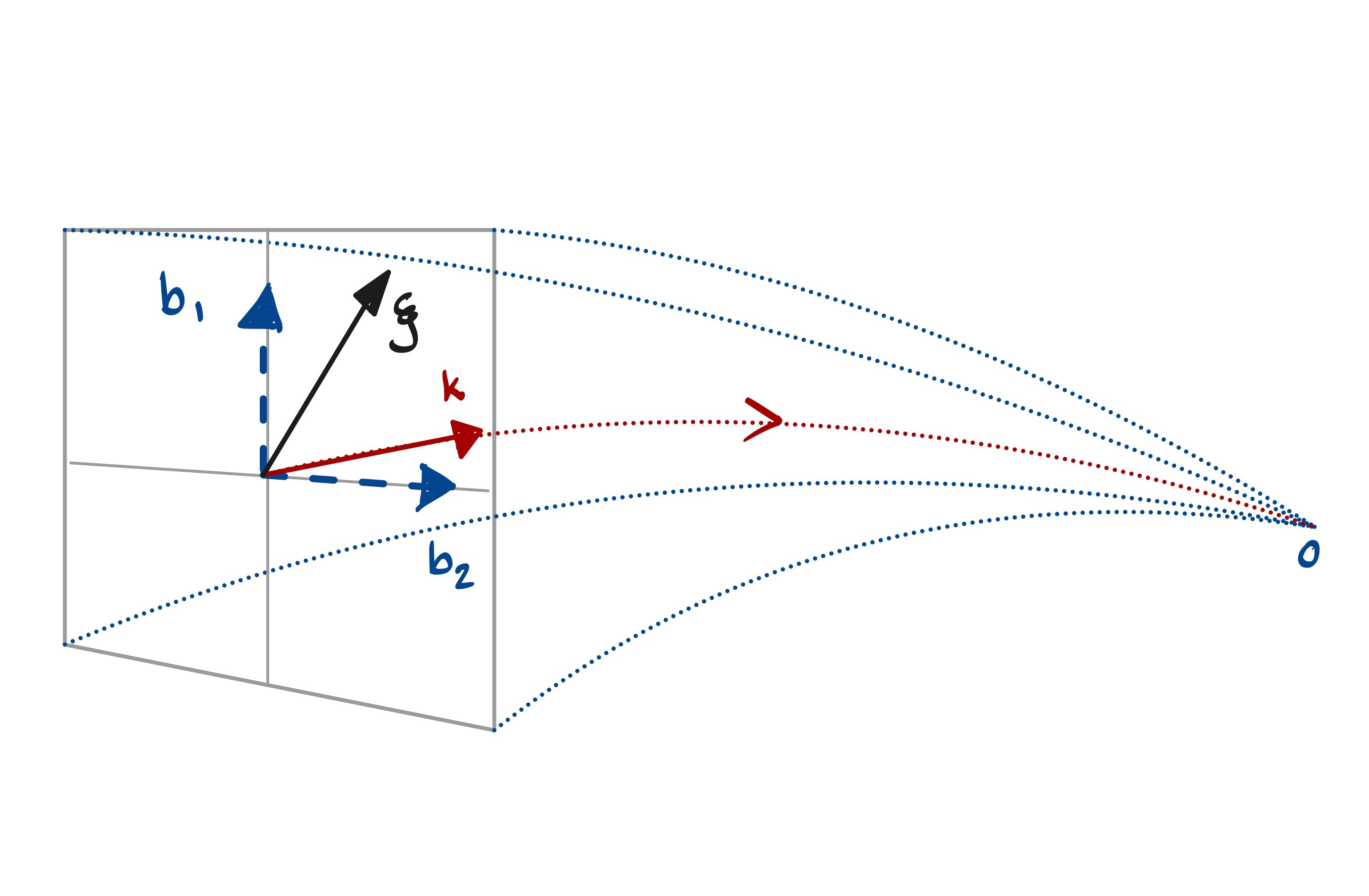}
    \caption{Screen space.}
    \label{fig:placeholder}
\end{figure}
Consider a bundle of null-geodesics emitted by an extended source and converging to the observer at time
$\tn.$
The observer sees the source  subtending    a solid angle  
$\delta \Omega _o$. The cross-sectional area of this bundle at the position of the source, $dS_o(t)$,  is equal to the projected surface area of the source at the time 
$t<\tn$ when the photons were emitted.  
The relation  $dS_o(t) \equiv  d_A^2(t) d\Omega_o$
defines  {\it angular diameter distance $ d_A$} of the source  as the {\it area distance}, measured from  the observer  standpoint, and evaluated at  the time  
 $t$ of emission of the photons. 

In order to evaluate  $d_A$ one needs to calculate how the separation vector $\xi$ between two infinitesimally close photons with the same affine parameter $\lambda$, 
 \textit{i.e.} the transversal size of the null geodesic beam,  changes while propagating in a generic spacetime from the source to the receiver.
To this purpose it is useful to decompose the separation 4-vector $\boldsymbol{\xi}$  into
the Sachs basis $\boldsymbol{b_I}$ (I=1,2) which span the 2D spatial plane  (vith metric $\delta_{IJ}$) perpendicular to the 
beam wave-vector $k^{\mu}(\lambda)$ and to the observers line-of-sight  $n^{\mu}$ ( $\boldsymbol{\xi}= \xi^{I} \boldsymbol{b_I}$).
Note that the screen basis vectors $\boldsymbol{b_I}(\lambda)$ are   parallel transported along the beam
$\nabla_{\boldsymbol{k}}(\boldsymbol{b_I}) = 0$
since we assume that the observers are geodesic. 

The components of the separation vector evolve according to the bi-dimensional equivalent of the geodesic deviation equation, i.e., the  Sachs vector equation,
\begin{equation}
 \frac{\dd^2 \xi_{I}}{\dd\lambda^2}=\tau_{IJ}\xi^J ,
 \label{sachsvec15}
 \end{equation}
where 
\begin{equation}
\tau_{IJ} \equiv  R_{\mu \nu \alpha \beta} (b_I)^{\mu}k^{\nu}k^{\alpha}(b_J)^{\beta},
\label{ott}
\end{equation}
is the {\it optical tidal tensor}, a $2 \times 2$ symmetric matrix that connects the evolution of the light bundle with the curvature of spacetime.

A  unique solution of (\ref{sachsvec15}) 
is singled out by  specifying the two initial conditions
\begin{eqnarray}
\xi^{I}_0  & \equiv & \xi^{I}(\lambda=0) ,\nonumber  \\
\dot{\xi}^{I}_0 & \equiv &   \left . \frac{\dd\xi^{I}}{\dd \lambda}\right |_{0}.
\end{eqnarray}
We assume that the null geodesic bundle converges at the freely falling terrestrial observer placed at $ \lambda = 0$. This condition  fixes 
$\xi^{I}_0=0$. The solution can therefore only depend on the initial rate $\dot{\xi}_0^{I}$. 
Since the solution of a linear differential equation depends linearly  on the initial conditions, one can write it as 
\begin{align}
\xi_I(\lambda)= \left . \mathcal{D}_{IJ} \frac{\dd\xi^{J}}{\dd\lambda} \right |_0 .
\end{align}
 One can
 thus recast the problem of determining the evolution of the separation vector $\xi_I$ as the following initial value problem for the unknown  Jacobi map ({\it Sachs equations}):
\begin{eqnarray}
\frac{\dd^2  \mathcal{D}_{IJ}(\lambda) }{\dd \lambda^2} & = & \tau_{IK} \mathcal{D}^{K}_{\phantom{K}J},  \nonumber \\
\left . \mathcal{D}_{IJ}\right|_0 & = & 0,  \nonumber \\
\left . \frac{\dd\mathcal{D}_{IJ}}{\dd\lambda} \right |_0 & = & \delta_{IJ}.
\label{sachs2}
\end{eqnarray}

We can choose the  affine parameter $\lambda$  to coincide with the local Euclidean distance in the observer’s rest frame, so that 
 the initial rate can locally be interpreted as the arrival angle of two rays
\begin{align}  
\theta^I= \left . \frac{\dd\xi^I}{\dd\lambda}\right |_{\lambda =0},
\end{align}
which converge at the position of the observer, \textit{i.e.} $\theta^I$
 denotes the angle on the celestial sphere between  two rays which were physically separated by $\boldsymbol{\xi}$ at an infinitesimal distance $\dd\lambda$ from the observer.

Integrating (\ref{sachs2}) 
 from the observer to a fiducial source located at a position corresponding to the affine parameter 
 $\lambda_s$ leads to
\begin{equation}
\xi_I(\lambda_s)=\left . \mathcal{D}_{IJ}(\lambda_s)\frac{\dd\xi^{J}}{\dd\lambda}\right |_0=\mathcal{D}_{IJ}(\lambda_s)\theta^J .
\label{uiui}
\end{equation}
Therefore the Jacobi matrix represents the transformation between angular coordinates (at the position of the terrestrial observer) into linear metric coordinates (at the source position).
Consequently,  the determinant  $\mathcal{D}$ is  the Jacobian of the coordinate transformation 
$\theta^I \rightarrow \xi^I$, \textit{i.e.} the ratio between the  infinitesimal `volume' at the source (the physical area of the source $\delta S_o$)  and
the `volume' at the observer position  (the observed solid angle $\delta \Omega_o$). Formally  
\begin{align}
 \left | \mathcal{D}(\lambda_s) \right | =\frac{\delta S_o ( \lambda_s)}{\delta \Omega_o}.   
\end{align}  
This relation can be  turned into an expression that allows us to estimate the  angular diameter distance,
\begin{equation}
d_A(z)= \sqrt{ \left | \mathcal{D}(\lambda_s) \right|}.
\label{jaco}
\end{equation}
The optical tidal tensor  $\tau_{IJ}$ can be decomposed into a pure-trace part and a trace-free part,
\begin{equation}
\tau_{IJ}=-\frac{1}{2}R_{\alpha \beta }k^{\alpha} k^{\beta}\delta_{IJ}+C_{\mu \nu \alpha \beta}(b_I)^{\mu}k^{\nu}k^{\alpha}(b_J)^{\beta} ,
\label{ott2}
\end{equation}
where $R_{\alpha \beta }$ is the Ricci tensor and 
\begin{eqnarray}
C_{\mu \nu \alpha \beta} = R_{\mu \nu \alpha \beta}+\frac{1}{2}\left[g_{\mu \beta} R_{\alpha \nu}- g_{\mu \alpha} R_{\beta \nu}+g_{\nu \alpha}R_{\beta \mu}-g_{\nu \beta}R_{\alpha \mu}\right]
+\frac{R}{6}\left[g_{\mu \alpha}g_{\beta \nu}-g_{\mu \beta}g_{\alpha \nu}\right],
\end{eqnarray}
is the Weyl tensor, 
which is endowed  with the same symmetries as the Riemann tensor $R_{\alpha \beta \mu \nu}$, and is furthermore
 trace-free,  $C^{\alpha}_{\phantom{\alpha} \beta \alpha \gamma}=0.$

Splitting (\ref{ott2}) into the Ricci and Weyl focusing terms 
allows us to better grasp the physical content of the geodesic deviation equation.  
The Ricci focussing originates from matter inside the null bundle and  causes 
$\boldsymbol{\xi}$  to increase or decrease homothetically. 
Note that if gravity is described by the Einstein equations,
\begin{align}
R_{\mu\nu}-\frac{1}{2}g_{\mu \nu}R+\Lambda g_{\mu \nu}=8\pi G  T_{\mu \nu},
\end{align}
then 
\begin{align}
\mathcal{R}=-4\pi G T_{\mu \nu}k^{\mu}k^{\nu}.
\end{align}
In the case of a pressureless perfect fluid with rest-frame energy
density $\rho_m$, the stress-energy tensor reads 
\begin{equation}
T_{\mu \nu} = \rho_m u_{\mu} u_{\nu} ,
\label{pfl}
\end{equation}
so that we get 
\begin{equation}
\mathcal{R} =  - 4 \pi G \rho_m (1+z)^2.
\label{riccif}
\end{equation}

A normal basis that satisfies the Sachs prescriptions is
\begin{equation}
b_{1}^{\mu}=\l(0,\alpha,\beta,0\r),\qquad b_{2}^{\mu}=\l(0,0,0,\gamma\r) ,\label{bcent}
\end{equation}
where 
\begin{align}
    \alpha &= -\beta \l(\frac{g_{\theta\theta}k^\theta+g_{r\theta}k^r}{g_{rr}k^r+g_{r\theta}k^\theta} \r),\\
    \beta &= \left[ \l(\frac{g_{\theta\theta}k^\theta+g_{r\theta}k^r}{g_{rr}k^r+g_{r\theta}k^\theta} \r)^2 g_{rr} +g_{\theta\theta} -2g_{r\theta}\l(\frac{g_{\theta\theta}k^\theta+g_{r\theta}k^r}{g_{rr}k^r+g_{r\theta}k^\theta} \r) \right]^{-\frac{1}{2}},\\
    \gamma &= \frac{1}{\sqrt{g_{\phi\phi}}}.
\end{align}
By using it, we obtain the expression of the Weyl focusing for the off-center observer, 
 \begin{align}
 C_{\mu \nu \alpha \beta}(b_I)^{\mu}k^{\nu}k^{\alpha}(b_J)^{\beta} = 
 \begin{pmatrix}
 W & 0  \\
0  &  -W 
\end{pmatrix}, \quad\textrm{where},
\end{align}

\begin{align}
    W&= \frac{r\l( rk^\theta - \Z k^r \sin{\theta} \r)^2 \l(  -a\KK_\textrm{q0}'+2a'\KK_\textrm{q0}+2 a^3\l( \HH_{_\textrm{q}}\HH_{_\textrm{q}}' + \dot\HH_{_\textrm{q}}'\r) \r)}{4 \l( ra' +a\l( 1+\Z\cos{\theta}\r)\r)}.
\end{align}
As a consequence, the optical matrix is diagonal and,  in an Szekeres cosmology where gravity is sourced by matter and $\Lambda$,  it is given by  
\begin{equation}\label{tidal tensor2}
\tau^I_{\phantom{I} J}=\left[-4\pi G \rho_m  \left(1+z\right)^2+ W\left(-1\right)^{I-1}\right] \delta^{I}_{\phantom{I}J}.
\end{equation}

Since the optical matrix  (\ref{tidal tensor2}) is diagonal, the  system (\ref{sachs2})  decouples  and the only non-trivial Sachs equations are 
\begin{eqnarray}
\frac{\dd^2}{\dd \lambda^2}\mathcal{D}_{11}(\lambda) & = & \tau_{11} \mathcal{D}_{11}, \nonumber \\
\frac{\dd^2}{\dd \lambda^2}\mathcal{D}_{22}(\lambda) & = & \tau_{22} \mathcal{D}_{22}, \nonumber \\
\mathcal{D}_{11}(0) & = & \mathcal{D}_{22} = 0, \nonumber \\
\left. \frac{\dd\mathcal{D}_{11}}{\dd\lambda} \right|_0 & = & \left.  \frac{\dd\mathcal{D}_{22}}{\dd\lambda} \right|_0 =1 .
\label{decsac}
\end{eqnarray}
Indeed, since the differential equations are homogeneous,  $\mathcal{D}_{12}(\lambda)=\mathcal{D}_{21}(\lambda)=0$ is the unique solution 
that satisfies the  given initial conditions.
The angular distance is thus 
 \begin{equation}
d_A(\lambda)=\sqrt{\mathcal{D}_{11}\mathcal{D}_{22}},
 \end{equation}
\textit{i.e.} the geometric mean of the diagonal terms of the Jacobian matrix $\mathcal{D}_{IJ}$.

\bibliographystyle{JHEP}
\bibliography{main}

\end{document}